\documentclass[a4paper,12pt,numbers=noenddot]{scrartcl}
\usepackage{authblk}
\usepackage[labelfont=bf,format=plain]{caption}
\usepackage{hep}
\usepackage{hyperref}
\usepackage{amsmath}
\usepackage{pslatex}
\usepackage{textcomp}
\usepackage{booktabs}
\usepackage{float}
\usepackage[page]{appendix}
\usepackage{multirow}
\usepackage{rotating}
\usepackage{rotfloat}
\usepackage{xspace}
\usepackage{cite}
\usepackage{subfig}
\usepackage{graphicx}
\graphicspath{{plots/}}

\usepackage{color}

\hypersetup{
  pdfpagemode={UseThumbs},
  pdfinfo={
    Author={P. Kant, O.M. Kind, T. Kintscher, T. Lohse, T. Martini,
      S. Moelbitz, P. Rieck, P. Uwer},
    Title={HATHOR for single top-quark production: updated predictions and
      uncertainty estimates for single top-quark production in
      hadronic collisions}
  }
}

\def\Fig#1{Fig.~\ref{#1}}
\def\Tab#1{Tab.~\ref{#1}}
\def\Eq#1{Eq.~(\ref{#1})}

\def\Ref#1{Ref.~\cite{#1}}
\def\Refs#1{Refs.~\cite{#1}}
\def\App#1{App.~\ref{#1}}
\def\SM{SM\xspace}
\newcommand{\Hathor}{\textsc{HatHor}\xspace}
\newcommand{\MCFM}{MCFM\xspace}
\newcommand{\mt}{\ensuremath{m_{\text{t}}}\xspace}

\newcommand{\Fsimple}[0]{\ensuremath{\tilde{f}}}
\newcommand{\F}[1]{\ensuremath{\tilde{f}^{(\mathrm{#1})}}}
\newcommand{\FF}[2]{\ensuremath{\tilde{f}^{(\mathrm{#1})}_{#2}}}

\newcommand{\f}[1]{\ensuremath{f^{(#1)}}}
\newcommand{\ff}[2]{\ensuremath{f^{(#1)}_{#2}}}

\newcommand{\PqqbV}[1]{\ensuremath{P^{V(#1)}_{\mathrm{q\bar{q}}}}}
\newcommand{\Pqq}[1]{\ensuremath{P^{(#1)}_{\mathrm{qq}}}}
\newcommand{\PqqV}[1]{\ensuremath{P^{V(#1)}_{\mathrm{qq}}}}
\newcommand{\PqqS}[1]{\ensuremath{P^{S(#1)}_{\mathrm{qq}}}}
\newcommand{\Pgq}[1]{\ensuremath{P^{(#1)}_{\mathrm{gq}}}}
\newcommand{\Pqg}[1]{\ensuremath{P^{(#1)}_{\mathrm{qg}}}}
\newcommand{\Pgg}[1]{\ensuremath{P^{(#1)}_{\mathrm{gg}}}}
\newcommand{\fihad}{\ensuremath{F_{i/\mathrm{h}}}}
\renewcommand{\fiHad}{\ensuremath{F_{i/\mathrm{h}_1}}}
\renewcommand{\fjHad}{\ensuremath{F_{j/\mathrm{h}_2}}}
\newcommand{\fiNoHad}{\ensuremath{F_{i}}}
\newcommand{\fjNoHad}{\ensuremath{F_{j}}}
\newcommand{\LR}{L_{\text{R}}}
\newcommand{\LM}{L_{\text{M}}}
\titlehead{\hfill HU-EP-14/22} 

\title{\Large\Hathor for single top-quark production: Updated predictions and
uncertainty estimates for single top-quark production in hadronic
collisions}

\author{\normalsize P.~Kant}
\author{\normalsize O.M.~Kind}
\author{\normalsize T.~Kintscher}
\author{\normalsize T.~Lohse}
\author{\normalsize T.~Martini}
\author{\normalsize S.~M\"olbitz}
\author{\normalsize P.~Rieck}
\author{\normalsize P.~Uwer}

\affil{\normalsize Humboldt-Universit\"at zu Berlin, Institut f\"ur Physik,
  Newtonstra{\ss}e~15, 12489~Berlin, Germany}

\date{\normalsize\today}

\begin{document}

\maketitle

\begin{abstract}
  \noindent We present updated predictions for single top-quark
  production in hadronic collisions. The analysis is based on
  next-to-leading order QCD calculations.  The input parameters are
  fixed to recent measurements. We compare different PDF sets and
  investigate the related uncertainties. The impact of uncalculated
  higher orders is estimated using an independent variation of the
  renormalisation and factorisation scale. The theoretical predictions
  are compared with recent measurements from Tevatron and LHC.
  Furthermore, the cross section measurements are used to estimate the
  top-quark mass.  To perform the analysis we extended the publicly
  available \Hathor program to single top-quark production. We thus
  provide a unified framework for the fast numerical evaluation of
  total cross sections for top-quark production, which may be used for
  example in Standard Model fits. For future extensions towards NNLO
  accuracy, we include already all scale dependent terms at NNLO.
  We briefly describe how to use the program and provide all
  required tools to repeat the aforementioned analysis.
\end{abstract}

\section{Introduction}
\label{sec:intro}

In high energetic hadron-hadron collisions, top quarks are produced
dominantly in pairs via the strong interaction, but also singly
through the weak interaction. Being of electroweak origin, the rates for
single top-quark production are evidently reduced compared to
top-quark pair production. With cross sections of about one third of
the respective cross sections for inclusive top-quark pair production,
still a significant number of single top-quark events is produced at
the Tevatron and the \LHC. The experimental analysis of single
top-quark production is, however, challenging owing to the complicated
event signature and sizeable backgrounds. Despite these complications,
single top-quark production is highly interesting for various reasons:
it provides a direct probe of the $\Wtb$ coupling which is sensitive to
many models of physics beyond the Standard Model (SM)
\cite{Tait:2000sh,Alwall:2006bx}. In the \SM, single top-quark
production allows a precise study of the $V\!-\!A$ structure of the
charged current interaction. In addition, it gives a direct handle to
the Cabibbo-Kobayashi-Maskawa (CKM) matrix element $\Vtb$ which is
otherwise only  measured indirectly. Furthermore, single top-quark
production provides a unique source of highly polarised top quarks and
may also offer valuable information to constrain the $\bquark$-quark
content in the proton.
\begin{figure}[h]
  \centering
  \subfloat[$t$-channel production]{%
    \begin{minipage}[b]{0.25\textwidth}
      \centering
      \includegraphics[width=1.\textwidth]{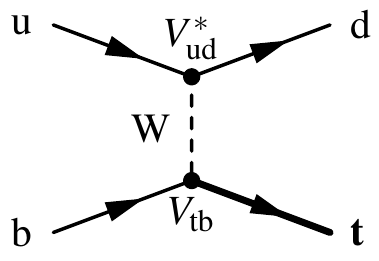} \\[-0.5ex]
      \label{fig:feynman-a}
    \end{minipage}
  } \hspace{0.03\textwidth}
  \subfloat[Associated $\Wboson\topquark$ production]{%
    \begin{minipage}[b]{0.26\textwidth}
      \centering
      \includegraphics[width=1.\textwidth]{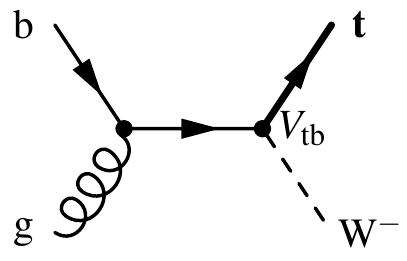} \\[-0.5ex]
      \label{fig:feynman-b}
    \end{minipage}
  } \hspace{0.03\textwidth}
  \subfloat[$s$-channel production]{%
    \begin{minipage}[b]{0.26\textwidth}
      \centering
      \includegraphics[width=1.\textwidth]{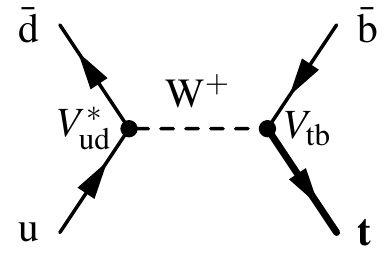} \\[-0.5ex]
      \label{fig:feynman-c}
    \end{minipage}
  } 
  \caption{Feynman diagrams in leading order for the dominant hard
    scattering processes in the three subprocesses of single top-quark
    production.}
  \label{fig:feynman}
\end{figure}

In \Fig{fig:feynman}, sample diagrams for single top-quark production
in the \SM are shown. The $t$-channel production dominates 
at both colliders, Tevatron and LHC. The next-to-leading
order (NLO) QCD corrections for the $t$-channel have been calculated
for the inclusive cross section in
\Refs{Bordes:1994ki,Stelzer:1997ns,Stelzer:1998ni}. In
\Refs{Harris:2002md,Sullivan:2004ie,Sullivan:2005ar} fully
differential results are presented, which are extended in
\Refs{Campbell:2004ch,Cao:2004ky,Cao:2005pq} to include also the
semi-leptonic decay of the top quark. In \Ref{Frixione:2005vw} the
fixed-order predictions are combined with parton shower results using
the MC@NLO framework. A similar study including also the $s$-channel
contributions using the POWHEG framework is presented in
\Ref{Alioli:2009je}.

At the LHC, the associated Wt production represents the second
important production channel; at the Tevatron this channel contributes
only at the level of a few per cent. The NLO corrections for the
associated production are given in fully differential form in
\Refs{Giele:1995kr,Zhu:2002uj}. At NLO, associated Wt production
interferes with leading order $\topquark\antitop$ production and the
subsequent decay $\antitop\to\Wboson^-\antib$.  The splitting
$\gluon\to\bquark\antib$ in the signal process prefers a small
transverse momentum $p^{\antib}_\topquark$ of the outgoing $\antib$.
This is not the case for the $\antib$ produced in the $\antitop$
decay.  One method to suppress the contribution from
$\topquark\antitop$ production and to disentangle to some extent the
two processes is thus to apply a cut on the transverse momentum of the
$\antib$ in the final state.  Typically, values lower than $25-50 $
\GeV\ are used for this cut \cite{White:2009yt}. Alternatively, one
might use diagram removal or diagram subtraction techniques to
disentangle top-quark pair production and single top-quark production
\cite{Tait:1999cf}.  For associated production, the semi-leptonic
top-quark decay has been considered in \Ref{Campbell:2005bb}.
Recently, the fixed order NLO corrections were also combined with the
parton shower.  Two implementations have been presented: one in the
MC@NLO framework \cite{Frixione:2008yi} and another in the POWHEG
framework \cite{Re:2010bp}.

The third channel, the $s$-channel, is the second important channel at
the Tevatron while it gives only a small contribution at the LHC.  The
NLO corrections for the inclusive $s$-channel production are given in
\Ref{Smith:1996ij}. Fully differential results have been published
together with the results for the $t$-channel in \Ref{Harris:2002md}.
Results including the top-quark decay are given in
\Refs{Campbell:2004ch,Cao:2004ap}.  

In \Refs{Campbell:2009ss,Campbell:2009gj}, $t$-channel single top-quark
production has been studied in the four-flavour scheme. In this
approach the $\bquark$-quark is not considered as part of the proton.
The leading order process is thus a two-to-three process. The naive
expectation is that the two schemes should give similar results as
long as effects of the finite $\bquark$-quark mass can be neglected
and the observables are not affected by $\ln(m_\bquark)$ terms. Indeed,
the authors of \Refs{Campbell:2009ss,Campbell:2009gj} find a
reasonable agreement between the two approaches.

Beyond fixed order in QCD, also the impact of large logarithmic
corrections due to soft gluon contributions have been studied in
detail. Being universal, these corrections can be partially resummed to
all orders in perturbation theory. Alternatively one may use the
universal terms to construct an ansatz for the yet uncalculated
next-to-next-to-leading oder (NNLO) corrections. Both approaches are
extensively discussed in the literature. For more details we refer to
\Refs{Mrenna:1997wp,Zhu:2002uj,Kidonakis:2006bu,Kidonakis:2007ej,%
  Kidonakis:2010ux,%
  Kidonakis:2010dk,Kidonakis:2011wy}.  Very recently partial NNLO
results for single top-quark production in the $t$-channel have been
presented \cite{Brucherseifer:2014ama}. More precisely, the NNLO vertex
corrections and the related real corrections are calculated.
Double-box topologies are not included.  In
\Ref{Assadsolimani:2014oga} the reduction to master integrals is
presented for the full set of two-loop corrections. Unfortunately, the
highly non-trivial two-loop master integrals are still unknown.

The vast amount of available literature documents a solid theoretical
understanding of single top-quark production. Despite the
non-negligible event rates, it took however some time to discover
single top-quark production in hadronic collisions. It{} was first
observed in $\ppbar$ collisions at the \Tevatron by the \CDF and
\Dzero collaborations\ \cite{Aaltonen:2009jj,Abazov:2009ii}. In $\pp$
collisions, single top-quark production was discovered in the
$t$-channel by the \ATLAS and \CMS collaborations for a centre-of-mass
energy of $\unit[7]{\TeV}$\ \cite{Aad:2012ux,Chatrchyan:2012ep}.  In
addition, evidence for the associated $\Wt$ production was reported in
\Refs{PhysRevLett.110.022003,Aad:2012dj} and later observed by the
\CMS collaboration with a significance of more than five standard
deviations \cite{CMS-PAS-TOP-12-040}.  For $s$-channel production,
so far only upper limits are given by the \LHC experiments\ 
\cite{CMS-PAS-TOP-13-009,ATLAS:2011aia}, while the Tevatron
experiments recently reported the observation of the $s$-channel\ 
\cite{CDF:2014uma}.

The aforementioned experimental results reflect just the beginning of
a growing activity. Refined experimental analysis in combination with
larger data samples will lead to a steadily increasing precision of
the experimental studies. To make optimal use of these results in the
context of precision tests of the SM and New Physics searches, a
detailed comparison with theoretical predictions is mandatory. For
differential distributions, publicly available tools exist to perform
such a comparison at NLO accuracy.  At the parton level, for example,
the {\it Monte Carlo for FeMtobarn processes} (MCFM)
\cite{Campbell:2004ch,Campbell:2005bb,Campbell:2009ss,Campbell:2010ff}
represents a convenient framework. For the $t$ and the $s$-channel
also the program ZTOP \cite{Sullivan:2004ie} can be used. The Monte
Carlo generators MC@NLO \cite{Frixione:2005vw,Frixione:2008yi} and
POWHEG \cite{Alioli:2009je,Re:2010bp} may be used to incorporate also
the effect of the parton shower. As a matter of fact, these tools can
also be used to calculate predictions for inclusive cross sections.
However, since they are designed for differential distributions, this
requires in general a phase space integration which is done
numerically and may lead to an increased runtime. Detailed
investigations of inclusive cross sections like for example studies of
renormalisation and factorisation scale dependencies or uncertainty
studies due to the parton distribution functions (PDF) may be limited
by the available computing resources.  One aim of this article is to
provide a publicly available tool dedicated to the calculation of
inclusive cross sections, similar to what has been done for top-quark
pair production
\cite{Aliev:2010zk,Ahrens:2011mw,Czakon:2011xx,Beneke:2012wb}.  Using
inclusive cross sections at the parton level as input for the
convolution with the PDFs, the numerical phase
space integration is avoided.  This approach is implemented in the
\Hathor program which is thus extended to also allow state of the art
theoretical predictions for hadronic single top-quark production.  As
an application, we investigate in detail the impact of PDF
uncertainties and study the independent variation of the
renormalisation and the factorisation scale.  
Future applications
could be fits of the $\bquark$-PDF or a global data analysis in the
context of the \SM.  In the next section we outline the theoretical
setup, give a short description of the program and discuss its
validation.  Section~\ref{sec:usage} outlines how to use the \Hathor
program. In Sec.~\ref{sec:app} the program is applied to study in
detail theoretical uncertainties due to factorisation and
renormalisation scale variations and parton distribution functions.
In addition, the theoretical predictions are compared with recent
measurements.  Section\ \ref{sec:summary} summarises the main results
of this article.

\section{Theoretical setup}
\label{sec:description}

Throughout this article, we restrict ourselves to the inclusive
cross section for single top-quark production. The top-quark
mass $\mt$ is renormalised in the on-shell scheme.  In the QCD
improved parton model, the inclusive cross section for single top-quark
production is  calculated according to
  \begin{equation}
    \label{eq:fact}
    \sigmaHad(s) = \sum_{i,j}\iint\!\!\dUp x_1
    \dUp x_2 \,\fiHad(x_1,\muF)\,\, \fjHad(x_2,\muF)\,\,
    \sigmaPart_{ij}(\sPart;\,\alphaS(\muR),\muF)\,,
\end{equation}
where $s$ is the hadronic centre-of-mass energy squared. The
factorised cross section for the hard scattering of two incoming
partons $i$ and $j$ with partonic centre-of-mass energy squared
$\sPart\!=\!x_1x_2s$ is described by $\sigmaPart_{ij}$. The
renormalisation and factorisation scales are denoted by $\muR$ and
$\muF$, respectively.  The parton distribution functions are given by
$\fihad(x,\muF)$. Qualitatively, they describe the probability to find
the parton $i$ inside the hadron h with momentum fraction between $x$
and $x+\dUp x$.  The QCD coupling $\alphaS(\muR)$ is determined in a
scheme with five massless flavours. Note in particular that the
$\bquark$-quark is treated as massless. In the five-flavour scheme, the
$\bquark$-quark in the initial state is described by the appropriate
$\bquark$-PDF.  For each production channel the factorised partonic
cross section can be expanded in
\begin{equation}
  a(\muR) = {\alphaS(\muR)\over 2\pi}.
\end{equation}
Up to NNLO accuracy the expansion reads:
\begin{equation}
  \sigmaPart_{ij}(\sPart;\,\alphaS(\muR),\muF) = 
  a^k(\muR) \sigmaPart_{ij}^{(0)}(\sPart)
  +a^{k+1}(\muR) \sigmaPart_{ij}^{(1)}(\sPart;\,\muR,\muF)
  +a^{k+2}(\muR) \sigmaPart_{ij}^{(2)}(\sPart;\,\muR,\muF)\,.
   \label{eq:sigmaexpand}
\end{equation}
For $s$ and $t$-channel production we have $k=0$, while for the associated
production $k$ is equal to 1. For the leading order cross sections 
$\sigmaPart_{ij}^{(0)}$, compact analytic expressions exist and the
application of \Eq{eq:fact} is straightforward.

\subsection{Partonic cross section at NLO}
At NLO accuracy, no compact analytic expressions for the inclusive
cross sections are available in
closed form. Complete NLO calculations have been published. However, 
these calculations typically involve numerical phase space integrations of
virtual and real corrections. The complete
code to perform these integrations for all three production channels has
been published for example in the \MCFM  program. 
In
principle, the program can also be used to evaluate 
the partonic cross sections for different partonic energies. In
practice, the extraction is however non-trivial due to the complicated
structure of one-loop calculations and the fact that this option is
not intended by the authors of MCFM. The NLO cross section 
$\sigmaPart_{ij}^{(1)}$ includes contributions from
virtual corrections $\sigma^V_{ij}$, 
real corrections $\sigma^R_{ij}$
and the factorisation of initial state singularities 
$\sigma^{\mathrm{fac}}_{ij}$. Schematically we may write
\begin{eqnarray}
  a^{k+1}(\muR)\sigmaPart_{ij}^{(1)} &=&
  \sigma^V_{ij} + \sigma^R_{ij} + \sigma^{\mathrm{fac}}_{ij}\\
  &=&\int \dUp R_n {\dUp \sigma^V_{ij}\over \dUp R_n}
  + \int \dUp R_{n+1} {\dUp \sigma^R_{ij}\over \dUp R_{n+1}}
  + \int \dUp x\dUp R_{n} {\dUp \sigma^{\mathrm{fac},x}_{ij}\over \dUp R_{n}},
\label{eq:genericNLO}
\end{eqnarray}
where $\dUp R_n$ denotes the $n$ particle phase space measure. Each contribution
contains soft and collinear singularities which cancel in the sum.
To achieve this cancellation, the
Catani-Seymour subtraction formalism
\cite{Catani:1996vz,Catani:2002hc} is employed in MCFM, i.e.
local counter terms are added (and subtracted)
such that each contribution is rendered finite and the total result
remains unchanged. Technically, this introduces additional terms in the
calculation which are difficult to identify in a generic NLO code.
Furthermore, the factorisation of initial state singularities but also
the subtraction formalism, introduce additional convolutions (indicated
by the additional $x$ integration in \Eq{eq:genericNLO}).  These
convolutions obscure further the correct identification of the
partonic centre-of-mass energy which is however required for the
calculation of the partonic cross sections.

To obtain the partonic cross sections at NLO accuracy, three
complementary approaches have been followed in this work.  In order to
avoid any intervention concerning the MCFM code, the first approach
was to run \MCFM using pseudo-PDFs which allow for an extraction of
the partonic cross section $\sigmaPart_{ij}^{(1)}$. For the partonic
channel of interest, we use narrow Gaussian distributions as parton
distribution functions to probe the partonic cross section
$\sigmaPart$ at the partonic centre-of-mass energy
$\sPart\!=\!x_0^2s$.  More precisely we set
\begin{equation}
  \label{eq:pseudoPDF}
  \fihad(x,\mu_0)= \frac{1}{\sqrt{2\pi}\delta}
  \exp{\left(-\frac{(x-x_0)^2}{2\delta^2}\right)},
\end{equation}
where $\delta$ describes the width of the Gaussian distribution.
Those PDFs leading to additional partonic channels are set to zero.
Using the pseudo-PDFs, we obtain to good approximation
\begin{equation}
  \left.{\sigmaHad}(s;\,\mu_0,\mu_0)\right|_{\mbox{\scriptsize pseudo-PDFs}} 
  = \sigmaPart_{i,j}(x_0^2s;\,\mu_0,\mu_0)+\mathcal{O}(\delta^2).
\end{equation}
The systematic error of the extracted cross section is proportional to
the square of the width  of the pseudo-PDFs. Further
uncertainties arise at the threshold when a part of the pseudo-PDF is
essentially cut off and the pseudo-PDF is thus no longer
normalized to 1. 
Within these limitations, a precise extraction of $\sigmaPart$ is
possible over a wide range of partonic centre-of-mass energies,
provided the width $\delta$ is chosen small enough. The self
consistency can be checked by using different values for $\delta$.
Note that the procedure is only applied once for $\muR=\muF=\mu_0$,
since the full renormalisation and factorisation scale (in)dependence can
be restored using the renormalisation group equation.

\begin{figure}[htbp]
  \centering
  \includegraphics[width=0.9\textwidth]{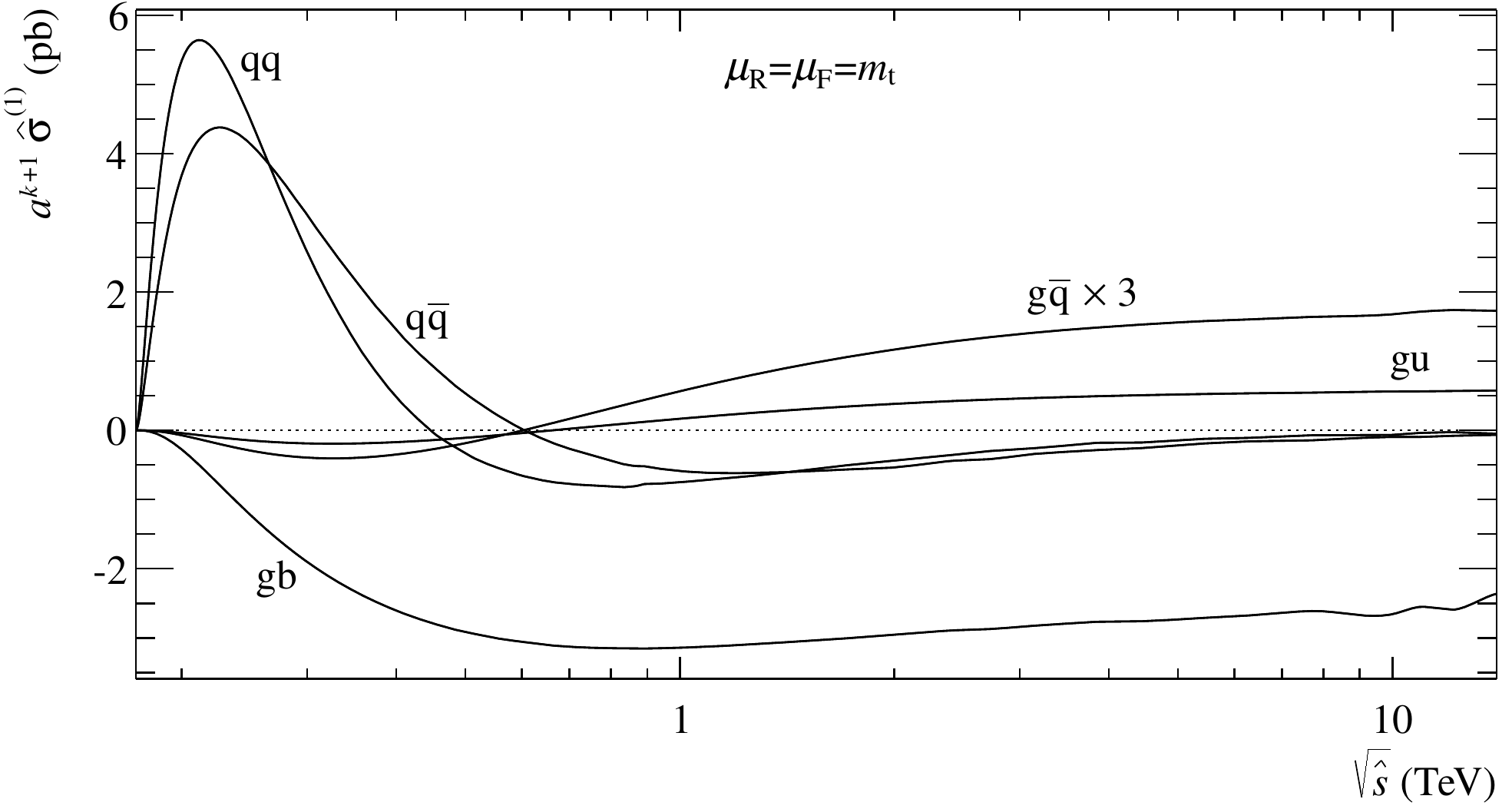}
  \caption{NLO contributions to the partonic cross section for single top-quark 
    production ($t$-channel) as a function of the partonic
    centre-of-mass energy for a top-quark mass of \unit[172.5]{\GeV}.
    Both scales $\muR$ and $\muF$ are set to the mass of the top
    quark. Here, $\quark$ and $\antiquark$ indicate all applicable
    flavours for the given channel, whereas $\upquark$ denotes an up
    or charm quark and $\bquark$ denotes a down-like quark.}
  \label{fig:partonicxs}
\end{figure}
This approach was validated by reproducing the known scaling functions
for top-quark pair production. In addition, we applied the method to
reproduce the known analytic results for single top-quark production
in leading-order.  Subsequently, the NLO corrections for single
top-quark production have been
extracted. Close to threshold, where small values of $\delta$ are
required, a significant numerical effort is necessary since MCFM is
not designed to cope with the employed type of PDFs. Despite this drawback, the
method is universally applicable to extract partonic cross sections from
generic NLO programs without the need to modify the
respective programs.

As an alternative approach we have studied the internal structure of
MCFM to identify the partonic cross sections. Once this was done, we
were able to remove the integration over the parton distribution
functions and restrict the numerical integration to the phase space
and convolution integrals.\footnote{A similar approach is followed in
  the APPLGRID framework\ \cite{Carli:2010rw}.} Evidently, this
approach is numerically much more efficient than using pseudo-PDFs and
leads to a shorter runtime and numerically much more precise partonic
cross sections. Furthermore, being able to remove the integration over
the PDFs analytically eliminates the systematic error associated with
the pseudo-PDFs. The downside of this approach is the necessity for
considerable changes of the MCFM program.

For a given value of the top-quark mass, the results of these two
approaches have been compared for all production channels and all
corresponding initial states. All results agree with each other within
less than 1\,\textperthousand, which proves the correctness of the two
methods.

As a third approach, we have written a private code implementing the
virtual corrections as given in \Ref{Harris:2002md}.
We have restricted ourselves to $s$
and $t$-channel production. Again we found perfect agreement with the
results extracted using the two methods described above. 
 
\begin{table}[ht!]
  \centering
  \begin{tabular}{llcc}
    \toprule
    Variable & Default value & Fixed & Valid range \\
    \midrule
    $\sqrt{s}$ & $\unit[1.96]{\TeV}$ ($\ppbar$), $\unit[8]{\TeV}$
    ($\pp$) 
    & no & any \\
    Colliding hadrons & $\pp$ & no & $\pp$, $\ppbar$ \\
    Top charge & $\topquark$  & no & $\topquark$, $\antitop$,
    $\topquark+\antitop$ \\
    $(\hbar{}c)^2$ & $\unit[3.89379323\!\times\!10^8]{\GeV^2pb}$ & no &
    --- \\
    $\sin^2\theta_{\text{W}}$ & $0.2228972$ & no & --- \\
    $\alpha(m^2_\Zboson)$  & $\nicefrac{1}{132.2332298}$ & no & ---\\
    $\alphaS$ & depends on chosen PDF & no & any appropriate PDF \\
    $m_\Wboson$ & $\unit[80.385]{\GeV}$ & yes & --- \\
    $m_\topquark$ (pole mass) & $(\unit[173.5]{\GeV})$ & no
    & 165\,--\,$\unit[950]{\GeV}$ \\
    $\muR$, $\muF$ & $(m_{\topquark})$ & no & any \\
    CKM matrix & PDG\,2012 & no & any \\
    $\Pt^{\antib}$ (assoc. Wt\,prod.) & $\unit[\leq25]{\GeV}$ &
    yes & --- \\
    PDF set & (CT10\,nlo) & no & any from \LHAPDF \\
    \bottomrule
  \end{tabular}
  \caption{Collection of  all constants and parameter settings of the
    \Hathor program. Their choice is based on latest world averages
    \cite{Beringer:1900zz}. Some of the model parameters can be varied
    by the user (cf.\ Sec.~\ref{sec:usage}), while the other values
    are fixed. Unless otherwise stated, the values in brackets are used
    throughout the figures in this paper, but they do not have defaults in
    \Hathor. All PDF sets available in the Les Houches Accord PDF
    Interface (\LHAPDF)\ \cite{Whalley:2005nh} can be used. The value
    for $\alphaS$ is predetermined by the chosen PDF set. Several PDF
    groups provide sets with varied $\alphaS$ values which can be used
    to study the $\alphaS$ dependence (see Fig.\ \ref{fig:alphasdep}).}
  \label{tab:settings}
\end{table}
In Fig.\ \ref{fig:partonicxs}, we show as an example the NLO
$t$-channel cross sections for a top-quark mass of $m_\topquark =
\unit[172.5]{\GeV}$.  The corresponding plots for $s$-channel and
$\Wboson\topquark$ production are given in \App{sec:app-addfigs}
(Fig.\ \ref{fig:app-partxsdep-a} and Fig.\ \ref{fig:app-partxsdep-b}).
Note that single top-quark production depends on a variety of input
parameters, in contrast to top-quark pair production. While in
top-quark pair production it is possible to factor out the top-quark
mass and $\alphaS$, and define dimensionless scaling functions which
depend only on ${4\mt^2/\hat s}$ --- allowing to keep the full
top-quark mass dependence --- this is not possible for single
top-quark production. In addition to the top-quark mass, single
top-quark production depends also on the $\Wboson$ mass.  The full
mass dependence can thus not be encoded in a one-dimensional function.
Furthermore, the cross section depends also on the weak coupling and
the CKM matrix elements. This dependence is, however, multiplicative
and can be factored out.  For the numerical extraction of the partonic
cross sections, we have used the present world average
\cite{Beringer:1900zz} for the respective input. In
\Tab{tab:settings}, we show the default values adopted in this work.
Since in most cases the dependence on the input parameters is kept in
analytic form, their values can be easily changed. An exception is the
dependence on the $\Wboson$ boson mass. Since the W mass is measured
very precisely, we have fixed the value to the current world average
value. Possible future changes are expected to be only of the order of
a few MeV which will not lead to relevant changes of the cross section
for single top-quark production.  In case of the top-quark mass, we
have repeated the extraction for different input values allowing to
change the mass later in a wide range. This does not only allow to
study the $\mt$ dependence of cross sections, but also to compute
cross sections for electroweak production of hypothetical heavier
versions of the top quark.  The partonic cross sections have thus been
sampled for 18 different values of the top-quark mass and 64 different
values of the partonic centre-of-mass energy. Since the energy
threshold for any given production process depends on $m_\topquark$,
the sampling points for the centre-of-mass energy have been chosen
individually for each value of the top-quark mass, starting with the
production threshold $M$. For the $s$ and $t$-channel, $M$ equals
$m_\topquark$, whereas in the case of $\Wboson\topquark$ production $M
= m_\Wboson+m_\topquark$. The majority of the sampling points is
placed close to the threshold and the spacing between adjacent points
increases with higher energies.  Therefore the resulting grid is not
equidistant in $\sqrt{\hat s}$.  In the calculation of the hadronic
cross section, the partonic cross sections are required for arbitrary
values of the partonic centre-of-mass energy between the threshold $M$
and the collider energy. To obtain the cross section for intermediate
masses and intermediate partonic centre-of-mass energies from the
values stored in the two dimensional grid, we use two subsequent
interpolations. If the mass does not correspond to one of the values
used in the calculation of the grid, we interpolate first in the mass
to find the cross section values for the required mass and
neighbouring centre-of-mass energies. This first interpolation is
polynomial for $\mt$ below $\unit[165]{\GeV}$ and logarithmic above
$\unit[165]{\GeV}$.  In a second (polynomial) interpolation in the
centre-of-mass energy the cross section for the required
centre-of-mass energy is determined. In fact, we do not interpolate in
$\hat s$ but in $M^2/\hat s$.

In order to obtain the total
cross section, no kinematic cuts are applied and the entire phase space
is included in the integration. However, there is one exception. In the
case of the associated production of a $\Wboson$ boson and a top
quark, the issue of interference with top-quark pair production arises
at NLO. For initial states consisting of either a pair of gluons or a
pair of quarks, contributions from $\ttbar$ production can lead to
the same final state. The \MCFM program allows to impose a cut
on the transverse momentum $\Pt^{\antib}$ of the outgoing $\bquark$-quark,
which suppresses the contribution from $\topquark\antitop$
production.
In concordance with current experimental settings, the suggestion given
in Ref.\ \cite{White:2009yt} is followed, and an upper limit of
$\Pt^{\antib}\!\leq\!\unit[25]{\GeV}$ is used.

\subsection{Renormalisation and factorisation scale dependence}
Having extracted the partonic cross sections at NLO for the
renormalisation and factorisation scales equal to the top-quark mass,
the scale dependent terms at NLO and NNLO can be derived from the
renormalisation group equation. Here, the NNLO scale
dependence is calculated as a first step of a possible future
extension towards NNLO accuracy (all the scale dependent terms
presented in the following are included in the Hathor library, which will be
discussed in the next section).  For the derivation of the scale
dependent terms it is convenient to introduce for each partonic
channel scaling functions $f_{ij}$ defined by
\begin{equation}
  \hat \sigma_{ij} = \rho \, f_{ij}(\rho)\,,
\end{equation}
with $\rho={M^2/\hat s}$ as before. The total hadronic cross
section is then given by
\begin{equation}
   \sigmaHad(s;\,\muF,\muR) =
  \frac{M^2}{s}\sum_{i,j}
  \fiHad \otimes\, \fjHad\, \otimes\, f_{ij}\,,
\end{equation}
where the convolution  $\otimes$ is defined through
\begin{equation}
(f \otimes g) (x) = 
\int \dUp y\dUp z \delta(x-yz)f(y)g(z) = 
\int_{x}^{1}\frac{\dUp y}{y} f(y) g\left({x\over y}\right)\,.
\end{equation}
Using the Dokshitzer-Gribov-Lipatov-Altarelli-Parisi evolution
equation of the parton distribution functions,
\begin{equation}
   \label{eq:dglap}
 \frac{\dUp\fiNoHad(x,\mu)}{\dUp \ln(\mu^2)} 
 = a(\mu) \sum_{j} P_{ij} \, \otimes \, \fjNoHad(x,\mu)\,,
\end{equation}
for $\mu=\muF=\muR$, together with the QCD beta function
\begin{equation}
\label{eq:BetaFunction}
  {\dUp a(\mu)\over \dUp \ln(\mu^2)} = -\beta(a) 
= -a^2 \left(\beta_0 + a \beta_1 +\ldots\right)\,, 
\end{equation}
with 
\begin{displaymath}
\beta_0 = {1\over 2}\left(11 - {2\over3} n_f\right),
\quad \beta_1 = {1\over 4}\left(102-{38\over 3}n_f\right)\,,  
\end{displaymath}
and
\begin{equation}
  a(\muF) = a(\muR) \bigg(1+\beta_0 \LR a(\muR) + 
  \big[\beta_1 \LR + \beta_0^2 \LR^2\big]
    a(\muR)^2
  +\ldots\bigg)\,,
\end{equation}
it is straightforward to derive the
general structure of the scale dependent terms:
\begin{eqnarray}
  \label{eq:sigmalo}
  {1\over \rho}\sigmaPart^{(0)}(\sPart;\muR,\muF)&=&\f{0}\,, \\
  \label{eq:sigmanlo}
  {1\over \rho}\sigmaPart^{(1)}(\sPart;\muR,\muF) &=& \f{10} 
  + \LM \f{11} + k \beta_0 \LR \f{0}\,, \\
  \label{eq:sigmannlo}
    {1\over \rho}\sigmaPart^{(2)}(\sPart;\muR,\muF) &=& \f{20}
    + \LM \f{21} + \LM^2 \f{22} +\; k \beta_1 \LR \f{0}\notag \\
    &+& (k + 1) \beta_0 \LR \left(\f{10} + \LM \f{11}\right)
      + \tfrac{1}{2}k(k+1) \beta_0^2 \LR^2 \f{0},
\end{eqnarray}
where
\begin{displaymath}
  \LM\!=\!\ln\left(\frac{\muF^2}{m_\topquark^2}\right) \quad\mbox{and}\quad
  \LR\!=\!\ln\left(\frac{\muR^2}{\muF^2}\right).
\end{displaymath}
For simplicity we have suppressed the parton indices. 
Note that the scaling functions $f^{(i)}$ do not depend on $\muR$ and
$\muF$. Using
\begin{equation}
  P_{ij} = P_{ij}^{(0)} + 4\pi\alpha_s  P_{ij}^{(1)} + {\cal O}(\alpha_s^2),
\end{equation}
the functions $\f{11}, \f{21}, \f{22}$
are given by
\begin{eqnarray}
  \label{eq:f11}
  \ff{11}{ij} & = & - \sum_{m\,\in \qUp,\bar{\qUp},\gUp}\left(
                  P_{mi}^{(0)}\otimes\ff{0}{mj} 
                + P_{mj}^{(0)}\otimes\ff{0}{im}\right)
                + k \beta_0 \ff{0}{ij} \,,
\end{eqnarray}
\begin{eqnarray}
  \label{eq:f21}
  \ff{21}{ij} & = & - \sum_{m\,\in \qUp,\bar{\qUp},\gUp}\left(
    P_{mi}^{(1)}\otimes\ff{0}{mj} 
    + P_{mj}^{(1)}\otimes\ff{0}{im}\right)
  + k \beta_1 \ff{0}{ij} \nonumber\\
  && - \sum_{m\,\in \qUp,\bar{\qUp},\gUp}\left(
    P_{mi}^{(0)}\otimes\ff{10}{mj} 
    + P_{mj}^{(0)}\otimes\ff{10}{im}\right)
  + (k + 1) \beta_0 \ff{10}{ij}\,, 
\end{eqnarray}
\begin{eqnarray}
  \label{eq:f22}
  2\,\ff{22}{ij} &= & - \sum_{m\,\in \qUp,\bar{\qUp},\gUp}\left(
                    P_{mi}^{(0)}\otimes\ff{11}{mj} 
                  + P_{mj}^{(0)}\otimes\ff{11}{im}\right)
                  + (k + 1) \beta_0 \ff{11}{ij} \,.
\end{eqnarray}

The equations (\ref{eq:f11})--(\ref{eq:f22}) are universally applicable
to any process. Within the five-flavour scheme, the sum over q is taken
over u,d,s,c,b. In single top-quark production, the
cross sections depend on the flavour through the CKM\ matrix
elements. Factoring out the CKM\ matrix elements, the number of
independent cross section evaluations can be greatly reduced. Since the light
flavours are not distinguished in the final state it is useful to define
\begin{equation}
  \Vsum{q} = \left\{ 
    \begin{array}{cl}
      \absV{qd}^2 + \absV{qs}^2 + \absV{qb}^2
        & \quark\in\{\upquark,\charmquark,\topquark\} \\
      \absV{uq}^2 + \absV{cq}^2
        & \quark\in\{\downquark,\strangequark,\bquark\} \\
    \end{array}
  \right. \,.
\end{equation}
Each combination $f_{ij}$ can be split into a CKM-independent part,
$\Fsimple_{ij}$, and a CKM dependent coefficient. The functions
$\Fsimple$ are generic and depend only on the kinematic structure of
the corresponding Feynman diagrams. In the $s$-channel, the
scale-independent functions in LO and NLO are defined by:
\begin{equation}
  \ff{0}{\quark\antiquark} 
    = \FF{0}{\upquark\antidown}\, \absV{\quark\antiquark}^2\,\Vsum{\topquark} ,
   \quad
    \ff{i}{\quark\antiquark} 
    = \FF{i}{\upquark\antidown}\, \absV{\quark\antiquark}^2\, \Vsum{\topquark}
    , \quad
    \ff{i}{\gluon\quark}  = \FF{i}{\gluon\upquark}\,
    \Vsum{\quark}\,\Vsum{\topquark} ,\quad i\in \{10,11\},
\end{equation}
where in the $\quark\antiquark$ case we sum over the initial states
$\upquark\antidown, \upquark\antistrange, \upquark\antib,
\charmquark\antidown, \charmquark\antistrange, \charmquark\antib$,
while in the $\gluon\quark$ case the sum is taken over
$\gluon\upquark, \gluon\charmquark, \gluon\antidown,
\gluon\antistrange, \gluon\antib$. Having factored out the flavour
dependence, the results for $\F{11}$ can be simplified to
\begin{equation}\label{eq:fscales}
  \FF{11}{\upquark\antidown} =
   -2 \,\Pqq{0} \,\otimes\,\FF{0}{\upquark\antidown}\,, \quad
   \FF{11}{\gluon\upquark} =
   - \Pqg{0} \,\otimes\, \FF{0}{\upquark\antidown}\,.
\end{equation}
For the $t$-channel, the dependence on the CKM matrix elements looks as
follows:
    \allowdisplaybreaks[4] 
    \begin{align}
    \label{eq:ftchan:0qq}
    \ff{i}{\quark\quark'}& 
    = \FF{i}{\bquark\upquark}\, \Vsum{\quark}{\absV{\topquark\quark'}}^2 \,,
     &\quark\in \{ &
      \upquark, \charmquark\} \,,\quark' \in
      \{\bquark,\strangequark,\downquark\},
      i \in \{0,10,11\},\\
    \label{eq:ftchan:0qqb}
    \ff{i}{\quark\antiquark} &
    = \FF{i}{\bquark\antidown}\, \Vsum{\antiquark}\,{\absV{\topquark\quark}}^2 \,,
     &\quark\in \{ &
      \bquark,       \strangequark,  \downquark\} \,,\antiquark \in
      \{\antib,\antistrange,\antidown\},
      \\
    \label{eq:ftchan:1gu}
    \ff{j}{\gluon\quark} & 
    = \FF{j}{\gluon\upquark}\, \Vsum{\quark}\, \Vsum{\topquark} \,,
     &\quark \in \{ &
      \upquark, \charmquark\} \,,j\in\{10,11\}, \\
    \label{eq:ftchan:1gb}
    \ff{j}{\gluon\quark}  
    &= \FF{j}{\gluon\bquark}\, (\Vsum{\upquark} + \Vsum{\charmquark})\,
      {\absV{\topquark\quark}}^2 \,,
     &\quark \in \{ &
      \downquark, \strangequark,\bquark\} \, \\
    \label{eq:ftchan:1gqb}
    \ff{j}{\gluon\antiquark}  
    &= \FF{j}{\gluon\antidown}\, \Vsum{\antiquark}\, \Vsum{\topquark} \,,
     &\antiquark \in \{ &
      \antidown,\antistrange, \antib\} \,.
  \end{align}
The expressions for $\Fsimple$ read
  \allowdisplaybreaks[4] 
  \begin{alignat}{2}
   \label{eq:fscalet:qq11}
   \FF{11}{\bquark\upquark} &=&
   -2&\,\Pqq{0} \,\otimes\,\FF{0}{\quark\quark}\,, \\
   \label{eq:fscalet:qqb11}
   \FF{11}{\bquark\antidown} &=&
   -2&\,\Pqq{0} \,\otimes\,\FF{0}{\quark\antiquark}\,, \\
   \label{eq:fscalet:gu11}
   \FF{11}{\gluon\upquark} &=&
   - &\Pqg{0} \,\otimes\, \FF{0}{\quark\quark}\,, \\
   \label{eq:fscalet:gqb11}
   \FF{11}{\gluon\antidown} &=&
   - &\Pqg{0} \,\otimes\, \FF{0}{\quark\antiquark}\,, \\
   \label{eq:fscalet:gb11}
   \FF{11}{\gluon\bquark} &=&
     & \FF{11}{\gluon\upquark} + \FF{11}{\gluon\antidown}\,.
  \end{alignat}

In case of associated production of a $\Wboson$ boson and a top quark,
the dependence on the CKM matrix can be written in the following way:
    \allowdisplaybreaks[4] 
    \begin{align}
    \label{eq:fwtchan:0gq}
    \ff{i}{\gluon\quark} &
    = \FF{i}{\gluon\bquark}\, {\absV{\topquark\quark}}^2 \,,
    & \quad\quark \in \{ &
      \downquark,\strangequark,\bquark \} \,,\,\, i \in \{0,10,11\}, \\
    \label{eq:fwtchan:1qq}
    \ff{j}{\quark\quark} &
    = 2 \FF{j}{\bquark\upquark}\, {\absV{\topquark\quark}}^2 \,, 
    & \quad\quark \in \{ &
    \downquark,\strangequark,\bquark \} \,,\,\,j\in\{10,11\}, \, 
     \,\\
    \ff{j}{\quark_1\quark_2} &
    = \FF{j}{\bquark\upquark}\, {\absV{\topquark\quark_2}}^2 \,, 
    & \quad\quark_1 \in \{ &
    \upquark,\downquark,\charmquark,\strangequark,\bquark,\antiup,\antidown,
    \anticharm,\antistrange,\antib \} \,, \,\, 
    \quark_2 \in \{\downquark,\strangequark,\bquark\} \,,\,\,q_1 \ne q_2 ,\\
    \label{eq:fwtchan:1gg}
    \ff{j}{\gluon\gluon} &
    = \FF{j}{\gluon\gluon}\, \Vsum{\topquark} \,. 
    \end{align} 
    The scale-dependent terms $\Fsimple$ for the associated production
    are given by \allowdisplaybreaks[4]
  \begin{alignat}{2}
   \label{eq:fscalewt:gq11}
   \FF{11}{\gluon\bquark} &=&
     & \left(\beta_0\,-\,\Pqq{0}\,-\,\Pgg{0}\right)
       \,\otimes\,\FF{0}{\gluon\quark} \,, \\
   \label{eq:fscalewt:gg11}
   \FF{11}{\gluon\gluon} &=&
   -2 &\, \Pqg{0}\,\otimes\,\FF{0}{\gluon\quark} \,, \\
   \label{eq:fscalewt:qq11}
   \FF{11}{\bquark\upquark} &=&
   -  &\, \Pgq{0}\,\otimes\,\FF{0}{\gluon\quark} \,.
 \end{alignat}

The functions \F{21} and \F{22} which correspond to each \F{20}
function in the NNLO case are collected in \App{sec:app-scalingNNLO}.

\subsection{Validation}
In Tab.\ \ref{tab:settings}, all parameter settings used by the \Hathor
program are listed. Some of the values are fixed and cannot be changed
by the user. Most of the SM parameters as well as the PDF sets,
however, can be varied as described in Sec.~\ref{sec:usage}. 

\begin{figure}[t!]
  \centering
  \includegraphics[width=0.98\textwidth]{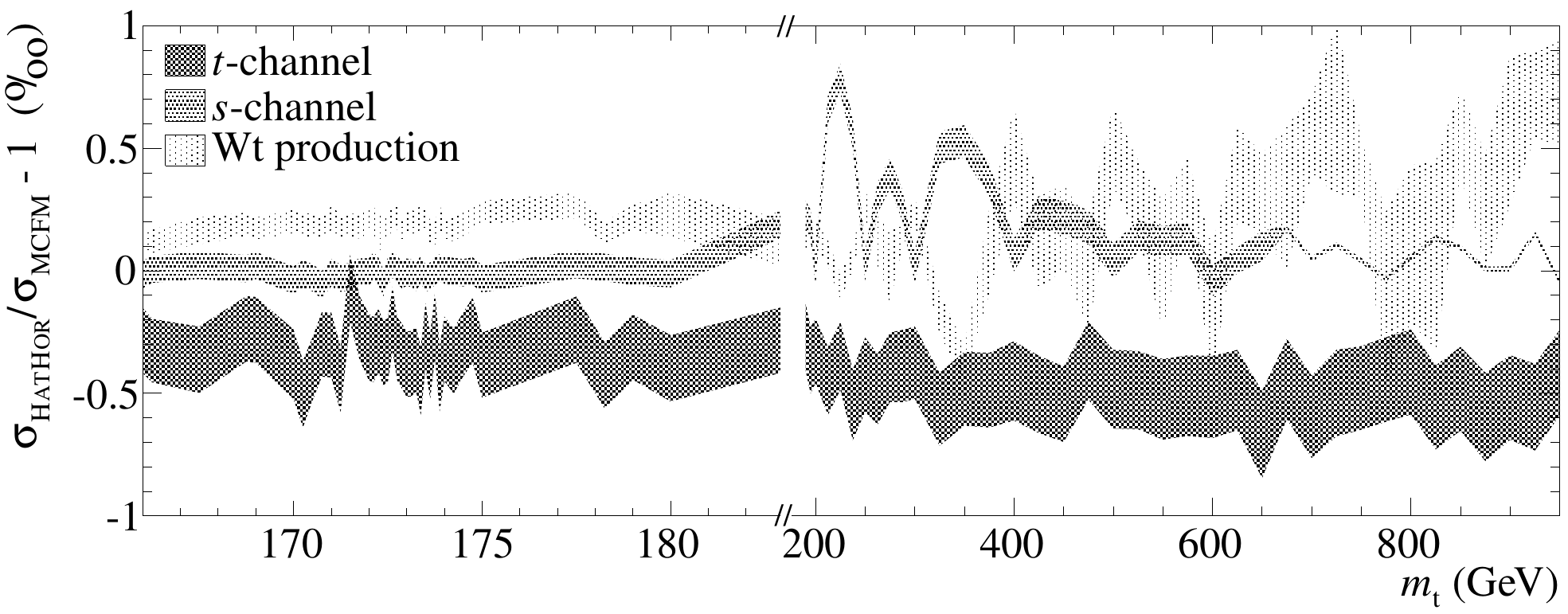}
  \caption{Cross sections for single top-quark
    production calculated with \Hathor for $\pp$ scattering at
    $\sqrt{s}\!=\!\unit[14]{\TeV}$, compared with the corresponding
    \MCFM results. For each production mode the relative deviation
    between both programs is shown. The width of the uncertainty bands
    indicates the statistical integration error of \MCFM. The
    interesting area around the actual top-quark mass is enlarged on
    the left side of the figure.}
  \label{fig:validation-nlo}
\end{figure}

The extension of the \Hathor program to single top-quark production
has been validated extensively for various parameter combinations.
The numerical implementation of the analytic results in LO was checked
with \MCFM. No significant deviation was found. The same was done with
the numerical computations in NLO. An example is shown in Fig.\ 
\ref{fig:validation-nlo}. Here, the relative deviation of the \Hathor
results from the corresponding \MCFM values for $t$-channel,
$s$-channel and associated $\Wboson\topquark$ production in $\pp$
collisions are shown as function of the top-quark mass.
The mass dependence is of interest since its computation with \Hathor
requires mass-dependent parameterisations of the partonic
cross sections and a proper interpolation. As can be seen from
Fig.~\ref{fig:validation-nlo}, 
the deviations are well below \unit[1]{\textperthousand} for
the whole mass range. In the vicinity of the measured top-quark mass,
the deviations are in most cases smaller 
than \unit[0.5]{\textperthousand}. 
Very similar results are observed for different centre-of-mass
energies and other parameter settings.

\section{\Hathor for single top-quark production}
\label{sec:usage}

This paper marks the release of the 2.0 version of the \Hathor
program, which is available for download at
\url{http://www.physik.hu-berlin.de/pep/tools}. 
The installation of
the package has changed with the new release compared to the previous 
versions. Detailed instructions how to install \Hathor can be found in
a release note distributed together with the library. 

Cross sections for top-quark pair production are evaluated in the same
way as before by using the
\texttt{Hathor} class, as described in\ \cite[Sec.\ 4]{Aliev:2010zk}.
For inclusive single top-quark production, the new classes
\texttt{HathorSgTopT}, \texttt{HathorSgTopS} and
\texttt{HathorSgTopWt} for the three production channels are
introduced. These classes and the \texttt{Hathor} class are derived
from a common base class and thus provide the same methods for the
calculation. Additionally, the following new common methods are
implemented in the new classes for single top-quark production:

\subsubsection*{\texttt{void getCkmMatrix(double ckm[3][3])}}
This method can be used to determine the CKM matrix currently used by
the \Hathor library. The values are copied into the array provided as
the argument.

\subsubsection*{\texttt{void setCkmMatrix(const double ckm[3][3])}}
This method allows to modify the CKM matrix. The default values are
set according to the recent PDG world averages \cite{Beringer:1900zz}.
To change the settings, a
$3\!\times\!3$ matrix with elements of type \texttt{double}
representing the magnitudes of the CKM matrix elements has to be
passed to this method. \Hathor does not check whether the provided
matrix is unitary. 

If only one entry of the CKM matrix needs to be
changed, a call of the {\tt getCkmMatrix} method may be useful:
\begin{verbatim}
Lhapdf pdf("CT10nlo");
HathorSgTopT hathor(pdf);
double ckm[3][3];
hathor.getCkmMatrix(ckm);
ckm[3][3]=1.;
hathor.setCkmMatrix(ckm);
\end{verbatim}

\subsubsection*{\texttt{PrintCkmMatrix()}}
This method prints the CKM matrix elements to the standard output.

\subsubsection*{\texttt{setParticle(PARTICLE particle)}}
This method allows the user to choose the charge of the top-quarks to
be produced. The three possible choices are defined as constants:
\texttt{TOPQUARK}, \texttt{ANTITOPQUARK} or
\texttt{BOTH}. The latter option evaluates the inclusive cross section
for single top-quark and anti-quark production. By default, the cross
section for single top-quark production is calculated. \\

The perturbative order of the cross section predictions can be
selected using the method \texttt{setScheme} as in the $\ttbar$ case:

\mbox{}\\\noindent\texttt{LO} \\
This option enables the leading order contribution, which is derived
from analytic results for the partonic cross sections implemented in \Hathor.

\mbox{}\\\noindent\texttt{NLO} \\
This option enables the next-to-leading order contribution, whose
partonic cross sections have been sampled from \MCFM. The
scale-dependent terms are calculated using the results presented in
the previous section. To obtain
the full total cross section in NLO the user has to combine this
option with the LO one.

The inclusion of the calculations into another program is demonstrated
as an example in the \texttt{demo\_sgtop.cxx} file . It makes use of
all features provided by \Hathor. Another example is given in
App.~\ref{sec:app-example}. In addition, we provide a graphical user
interface to the library, which may be useful if only a few
reference cross sections are required.

When publishing results obtained with the \Hathor program please quote
the references of the underlying calculations. At next-to-leading
order this is\ \cite{Campbell:2004ch} for $t$-channel and $s$-channel
production and\ \cite{Campbell:2005bb} for associated
$\Wboson\topquark$ production. 

\section{Results}
\label{sec:app}
A key feature of the \Hathor program is the possibility to vary
certain model parameters which allows fast computations of the total
cross section dependencies on such parameters. In this section, several
examples for such studies are presented. \\

\begin{figure}[t!]
  \centering
    \includegraphics[width=0.8\textwidth]{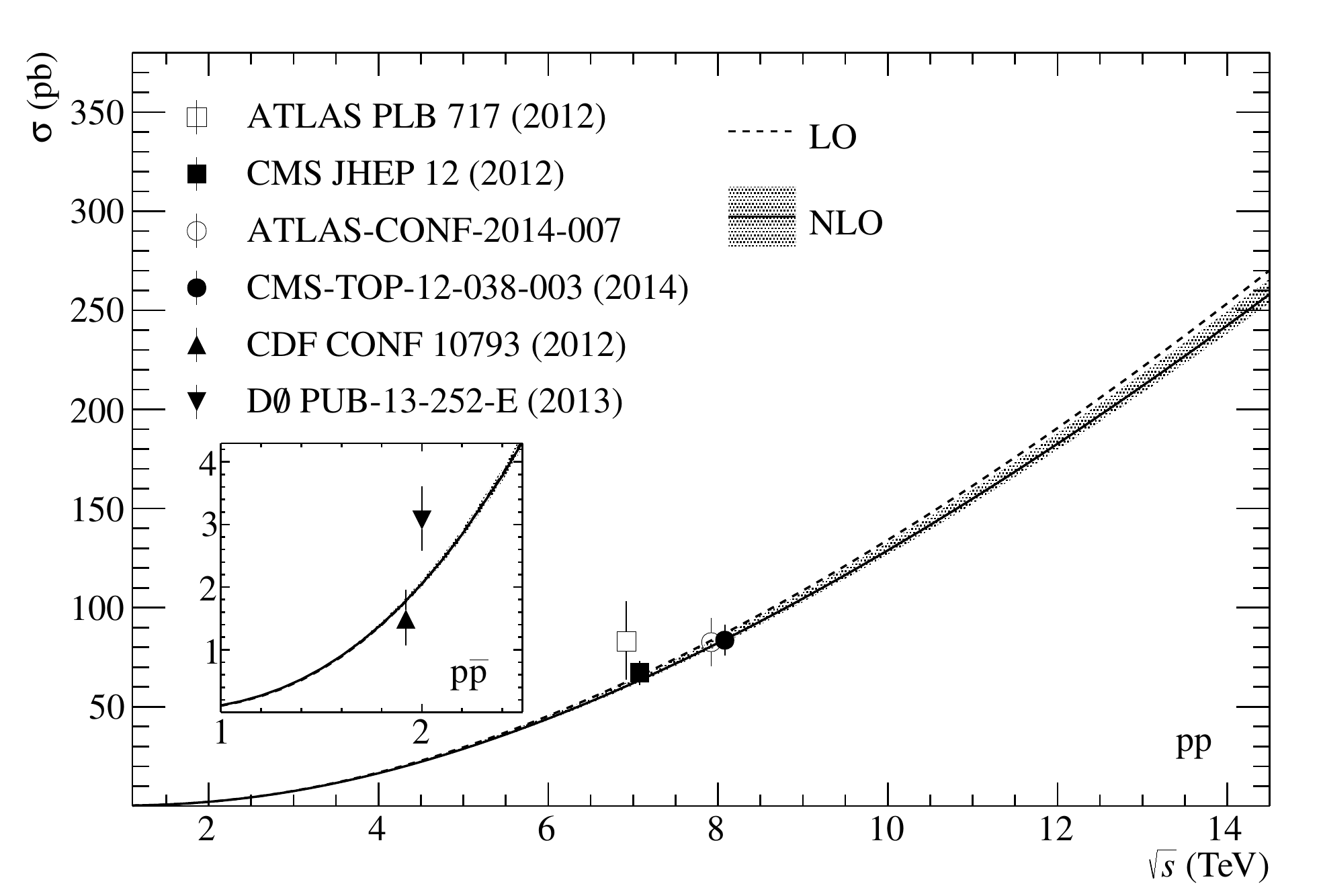}
    \caption{LO and NLO cross section for $t$-channel single top-quark 
      production calculated with the \Hathor program using CT10nlo as
      PDF set. The uncertainty band
      shown for NLO indicates the scale uncertainty.
      The latest measurements from the \LHC\ 
      \cite{Aad:2012ux,Chatrchyan:2012ep,ATLAS-CONF-2014-007,Khachatryan:2014iya}
      and the \Tevatron\ \cite{CDF:2013,D0:2013} are shown for
      comparison.}
  \label{fig:theouncert-t}
\end{figure}

\noindent\textit{Cross sections in higher orders} \\
Figure~\ref{fig:theouncert-t} shows the cross section for single top-quark
$t$-channel production as function of the centre-of-mass-energy in $\pp$
and $\ppbar$ scattering. Leading order and NLO calculations are
presented. The uncertainty bands indicate the uncertainty estimated by
the combined variation of $\muR$ and $\muF$ by factors between
$\nicefrac{1}{2}$ and 2. More detailed studies of the scale
dependence of the cross sections and their respective uncertainties
are given in the next paragraph. For comparison, also the latest measurements
from the \CDF and the \Dzero experiments as well as from \ATLAS
and \CMS are shown in \Fig{fig:theouncert-t}. The cross sections for the
$s$-channel and associated $\Wboson\topquark$ production are depicted
in Fig.\ \ref{fig:theouncert-sWt} in the appendix. On a standard 2.4
GHz Xeon CPU, the entire calculations for Fig. \ref{fig:theouncert-t}
were performed in about 15 minutes with the \Hathor program. \\

\begin{figure}[t!]
  \centering
  \subfloat[$t$-channel production]{%
    \includegraphics[width=0.48\textwidth]{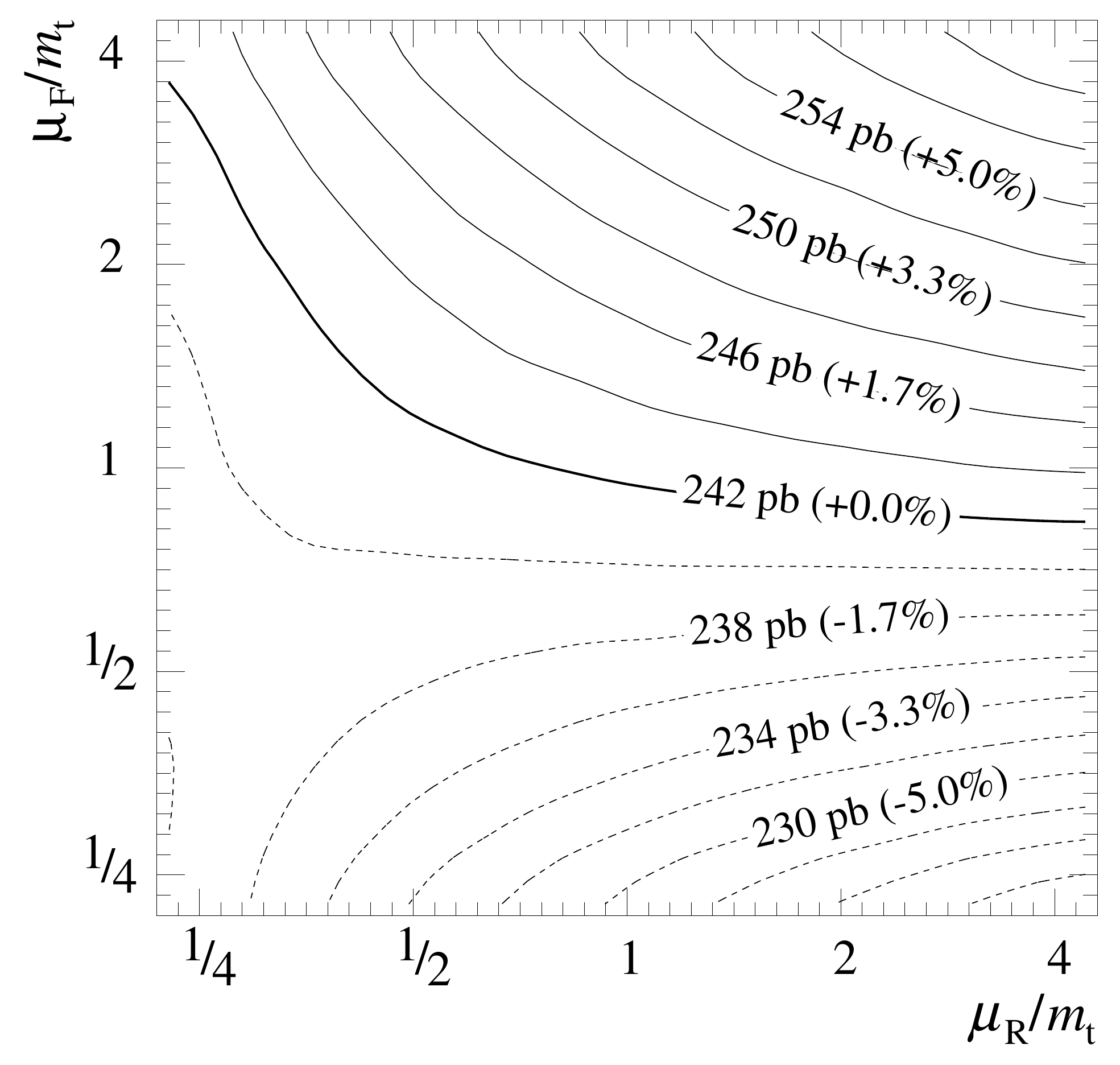}
    \label{fig:scales2d-a}
  } 
  \subfloat[$s$-channel production]{%
    \includegraphics[width=0.48\textwidth]{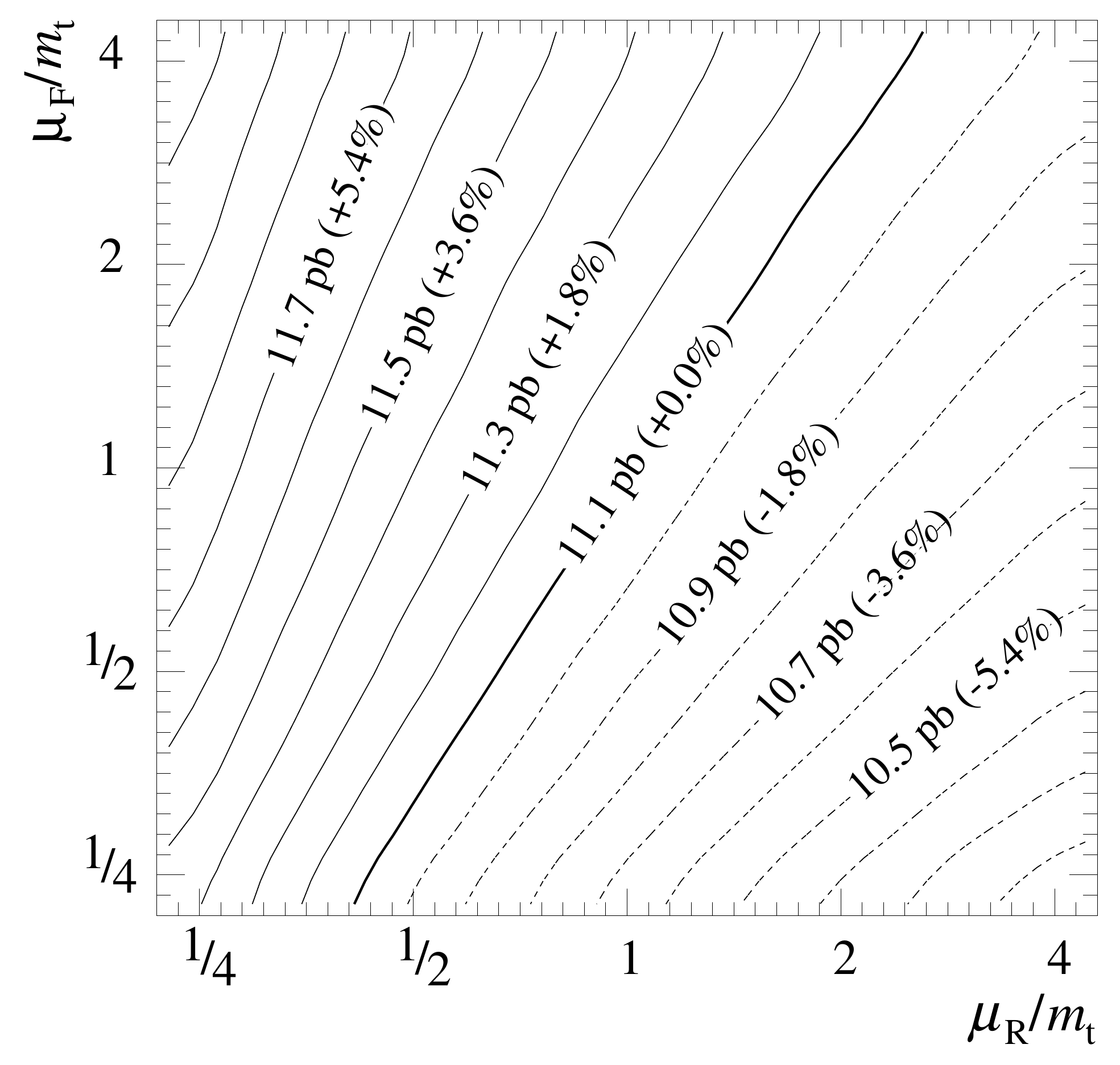}
    \label{fig:scales2d-b}
  } \\
  \subfloat[Associated $\Wboson\topquark$ production]{%
    \includegraphics[width=0.48\textwidth]{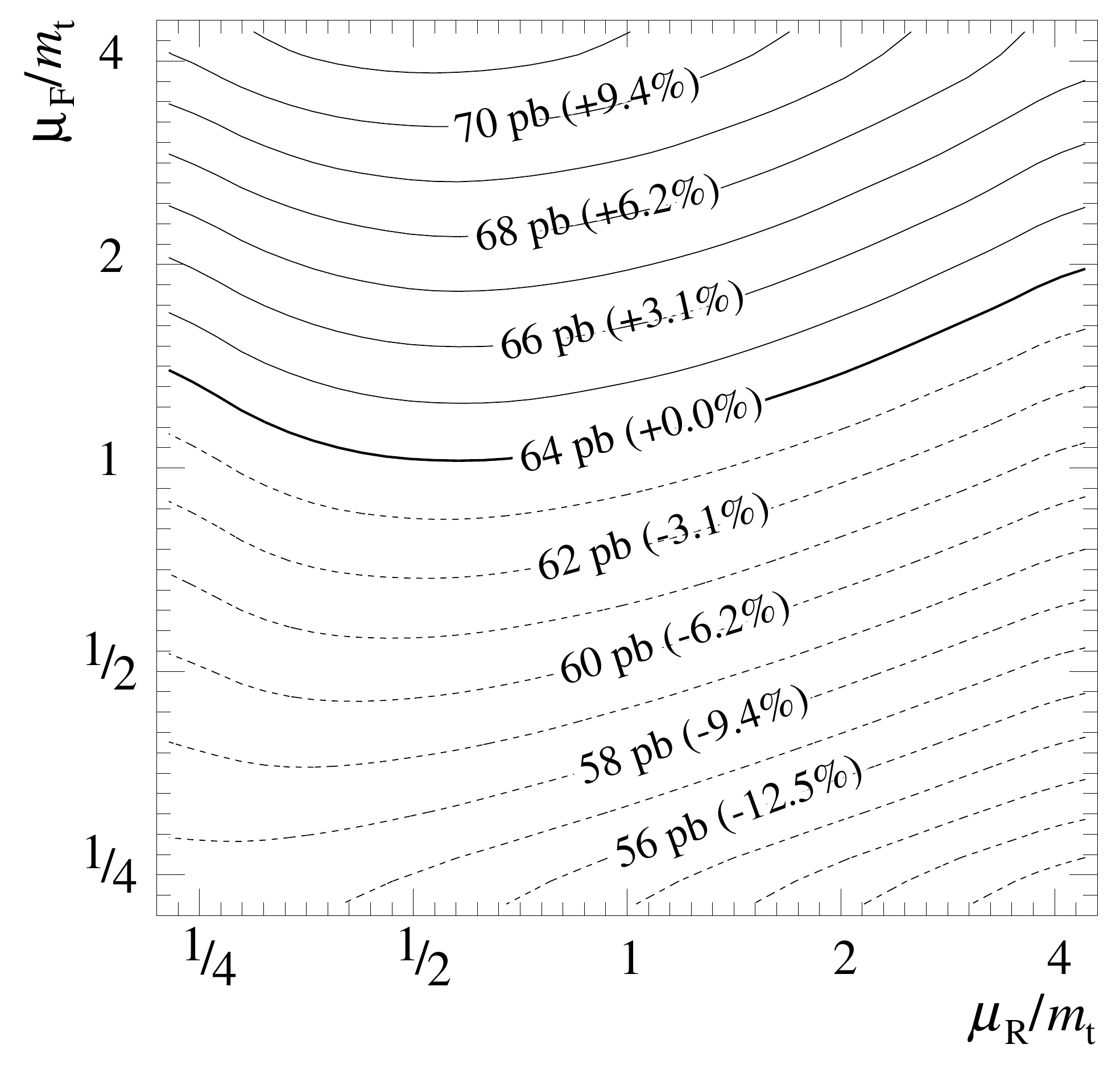}
    \label{fig:scales2d-c}
  }
  \caption{Two-dimensional cross section dependence on the
    renormalisation scale, $\muR$, and the factorisation scale,
    $\muF$, of the three single top-quark production channels in $\pp$
    collisions at $\sqrt{s}\!=\!\unit[14]{\TeV}$ in NLO using CT10nlo. The fat
    contour line in each figure indicates the reference cross section value for
    $\muR\!=\!\muF\!=\!\mt$. 
    Contour lines for cross sections below (above) the reference cross
    section are drawn as dashed (solid thin) lines.}
  \label{fig:scales2d}
\end{figure}

\noindent\textit{Renormalisation and factorisation scale (in)dependence} \\
The scale dependence of the cross section for the different orders is
given by Eqs.~(\ref{eq:sigmalo}--\ref{eq:sigmanlo}) which allow fast
computations for any given pair of $\muR$ and $\muF$. 
As an example, we show in \Fig{fig:scales2d} cross section contours
obtained from an independent variation of $\muF$ and $\muR$. 
Since $f_{ij}^{(20)}$ is not yet fully known, we include only the scale
dependence at NLO accuracy (i.e. Eq.~(\ref{eq:sigmanlo})) in the 
analysis.\footnote{Using an
  educated guess for the yet uncalculated $f_{ij}^{(20)}$ one may use
  Eq.~(\ref{eq:sigmannlo}) to estimate the possible improvements once the full
  NNLO calculation is available.}
In case
of $t$ and $s$-channel production, one observes that for a reasonable
change of the scales the cross section varies by less than five per
cent. In the $s$-channel the $\muF$ and $\muR$ dependencies are
anti-correlated.
Using a naive variation $\mt/2<\mu<2 \mt$ with $\mu=\muF=\muR$ thus
underestimates the scale uncertainty. In the $\Wt$-channel the residual
scale uncertainty is slightly larger. Again, a variation along the
diagonal in the $\muF-\muR$ plane underestimates the uncertainty.\\

\begin{figure}[t]
  \centering
  \includegraphics[width=0.8\textwidth]{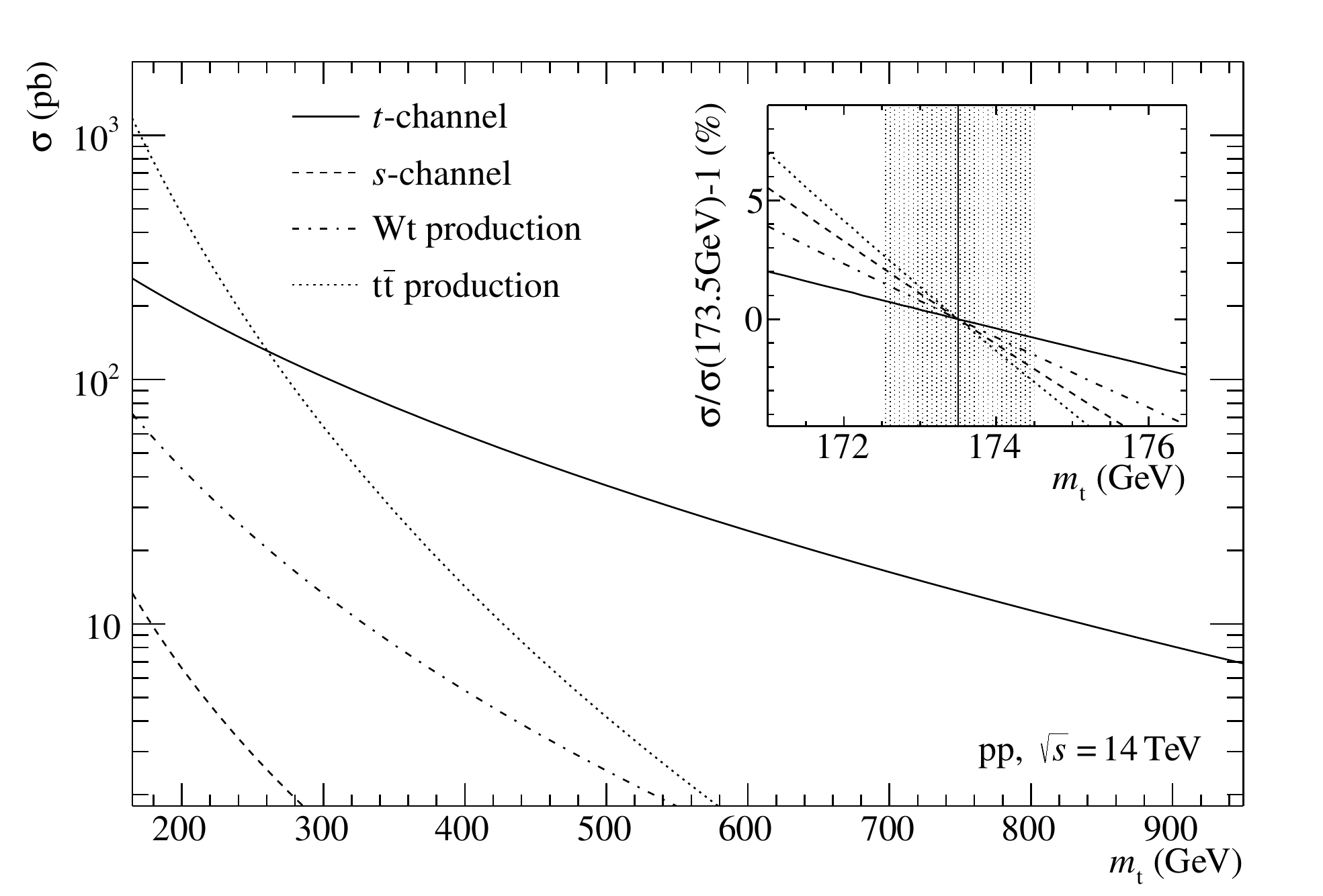}
  \caption{
    Top-quark production cross sections for different top-quark
    masses. The single top-quark production is calculated in NLO with
    CT10nlo, whereas top-quark pair production employs NNLO with
    CT10nnlo.  The in-set shows the relative change of the cross
    section around the actual top-quark mass.}
  \label{fig:massdep}
\end{figure}

\noindent\textit{Top-quark mass} \\
The \Hathor program also allows to vary the pole mass of the top
quark, $m_{\topquark}$, over a wide range. This is shown in Fig.\ 
\ref{fig:massdep} in NLO for the case of $\pp$ scattering at
$\unit[14]{\TeV}$. In the figure, the mass range around the measured
top-quark mass is shown together with a mass range extended up to
$\unit[1]{\TeV}$ which might be of interest for certain heavy quark
searches.

\begin{figure}[t!]
  \centering
  \includegraphics[width=0.94\textwidth]{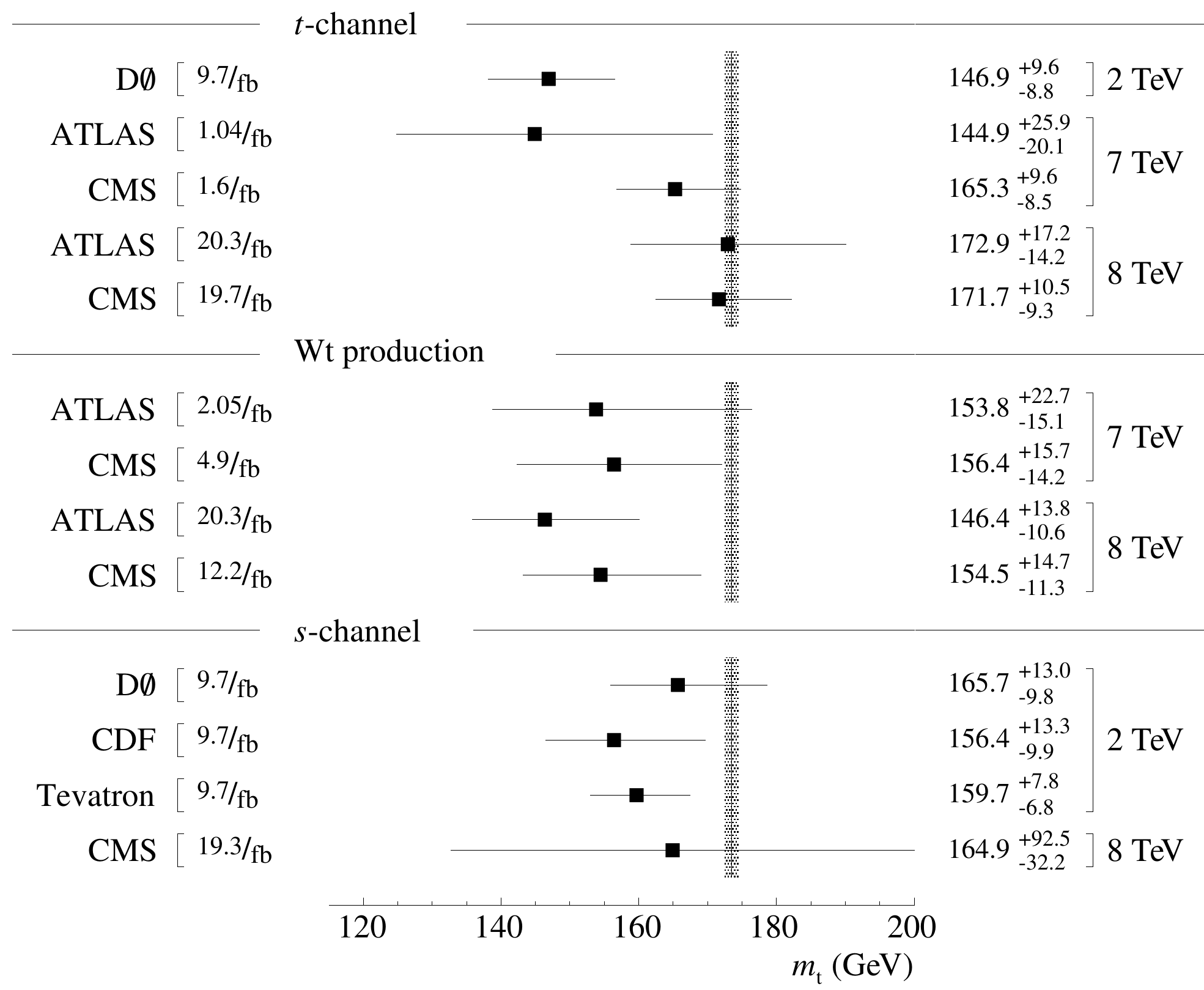}
  \caption{Overview of top-quark masses obtained from recent
    cross section measurements of single top-quark production.  All
    masses are given with respect to the NLO prediction using CT10nlo.
    Their numerical values are listed on the right side of the figure.
    All numbers are in $\unit{\GeV}$.  The vertical line denotes the
    current world average of the top-quark mass from measurements of
    $\ttbar$ production \cite{Beringer:1900zz}; the shaded band
    indicates the uncertainty.  The cross section values used here are
    taken from the measurements\ 
    \cite{Aad:2012ux,ATLAS-CONF-2014-007,Aad:2012dj,ATLAS-CONF-2013-100}
    of the \ATLAS collaboration, \cite{Aaltonen:2014qja} of the \CDF
    collaboration,
    \cite{Chatrchyan:2012ep,Khachatryan:2014iya,PhysRevLett.110.022003,%
      CMS-PAS-TOP-12-040,CMS-PAS-TOP-13-009} of the \CMS collaboration
    and\ \cite{D0:2013} of the \Dzero collaboration. The combined
    \Tevatron measurement is taken from\ \cite{CDF:2014uma}.}
  \label{fig:mass_measurements}
\end{figure}

Following up on \cite{Campbell:2009gj}, the cross section dependence
on the top-quark mass can be approximated to a precision of
$\unit[1]{\%}$ by
\begin{equation}
  \sigma(\mt) = \sigma(\bar{\mt})\left[1\,+\,
  A\times\left(\dfrac{\mt - \bar{\mt}}{\bar{\mt}}\right)\,+\,
  B\times\left(\dfrac{\mt - \bar{\mt}}{\bar{\mt}}\right)^2\right] \, ,
 \label{eq:mtapprox}
\end{equation}
where $\bar{\mt}$ is chosen to be $\unit[173.5]{\GeV}$\ 
\cite{Beringer:1900zz}. The coefficients for all three production
channels are given in Tab.\ \ref{tab:mtcoeff} in
App.~\ref{sec:app-coeff} for centre-of-mass energies at
$\unit[8]{\TeV}$ and $\unit[14]{\TeV}$, as well as for four different PDF
sets. 
The parameter $A$ relates the relative change of the cross section
with the relative change in $\mt$ close to $\bar{\mt}$.
For the three modes
of single top-quark production (not $\antitop$), the results for a
centre-of-mass energy of $\unit[8]{\TeV}$ are 
\begin{equation}
  \frac{\Delta\sigma_{t}}{\sigma_{t}} = -1.6\times
  \frac{\Delta\mt}{\mt}\, 
  ,\quad
  \frac{\Delta\sigma_{s}}{\sigma_{s}} = -3.9\times
  \frac{\Delta\mt}{\mt}\, 
  ,\quad
  \frac{\Delta\sigma_{\Wboson\topquark}}{\sigma_{\Wboson\topquark}} 
  = -3.1\times \frac{\Delta\mt}{\mt}\, ,
  \label{eq:xsecmt8}
\end{equation}
with slightly smaller slopes for an energy of $\unit[14]{\TeV}$,
\begin{equation}
  \frac{\Delta\sigma_{t}}{\sigma_{t}} = -1.4\times
  \frac{\Delta\mt}{\mt}\, 
  ,\quad
  \frac{\Delta\sigma_{s}}{\sigma_{s}} = -3.7\times
  \frac{\Delta\mt}{\mt}\, 
  ,\quad
  \frac{\Delta\sigma_{\Wboson\topquark}}{\sigma_{\Wboson\topquark}} 
  = -2.7\times \frac{\Delta\mt}{\mt}\, .
  \label{eq:xsecmt14}
\end{equation}
The parameter $A$ hence determines the experimental sensitivity for
extracting the top-quark mass from a measurement of single
top-quark production cross sections.
The production in the
$s$-channel exhibits the strongest top-quark mass dependence, in
particular for lower centre-of-mass energies (cf.\ Tab.\
\ref{tab:mtcoeff}). In
\Fig{fig:mass_measurements}, top-quark masses extracted
in this way from recent measurements are presented. The results are
given using the relation between the mass and the cross section at NLO
accuracy. Evidently, the uncertainty of the extracted top-quark masses
is large as a consequence of the weak sensitivity. Within the
uncertainties, the results are consistent with the world average.\\

\begin{figure}[t!]
  \centering
  \includegraphics[width=0.75\textwidth]{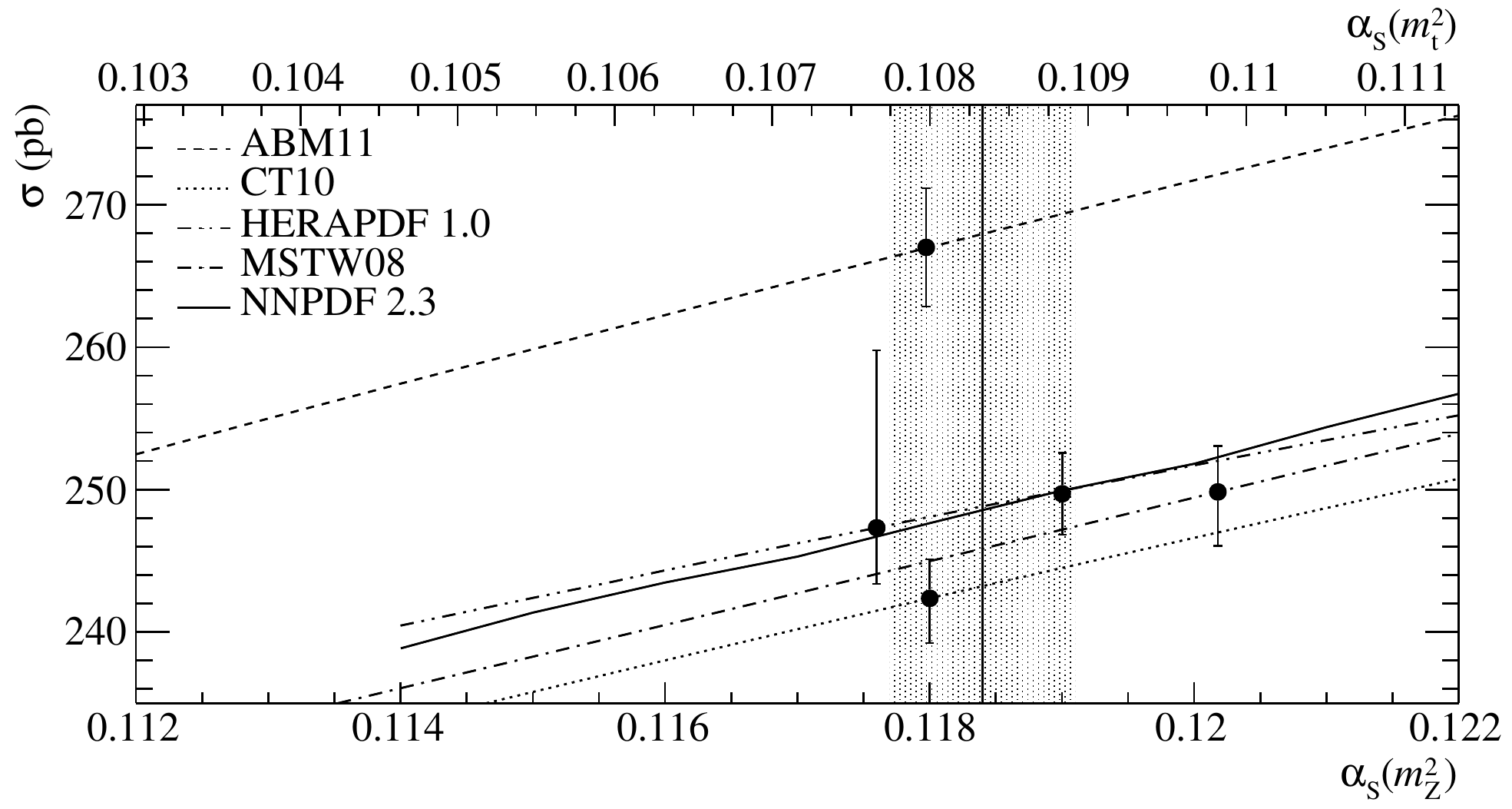}
  \caption{Single top-quark $t$-channel cross section in NLO for different
    values of $\alphaS$, computed for different NLO PDFs at
    $\sqrt{s}\!=\!\unit[14]{\TeV}$ in $\pp$ scattering. For each PDF
    set, the best fit value and the corresponding full PDF uncertainty
    is indicated by the black marker and the error bar. The $\alphaS$
    values are given with respect to the mass of the $\Zboson$ boson
    (lower abscissa) and the mass of the top quark (upper abscissa).
    The vertical line and the shaded box represent the latest world
    average value of $\alphaS$ and its uncertainty at the $\Zboson$ pole\
    \cite{Beringer:1900zz}.}
  \label{fig:alphasdep}
\end{figure}

\noindent\textit{Strong coupling constant}\\
The strong coupling constant, $\alphaS$, enters the computation
through the matrix elements and through the parton distribution functions.
Since the parton distribution functions and $\alphaS$ are highly
correlated when determined in global fits, it is not sufficient for
$\alphaS$ uncertainty studies to change only $\alphaS$. For a precise
estimate of the uncertainty, it is important to use PDF sets
providing fits for different $\alphaS$ values.  In \Fig{fig:alphasdep},
we show as an example the dependence of the single top-quark
$t$-channel cross section. The computation is performed at NLO
accuracy at a centre-of-mass energy of $\unit[14]{\TeV}$ for
proton-proton scattering. The best fit values for each PDF set as well
as the latest measurement for $\alphaS(m_{\Zboson}^2)$ are shown for
comparison. As one would have naively expected, we observe to good
approximation a linear dependence on $\alphaS$ with a slope close to
one. A one per cent uncertainty in $\alphaS$ thus roughly
translates to a one per cent uncertainty of the cross section. In
addition, \Fig{fig:alphasdep} shows that the ABM11 PDF
\cite{Alekhin:2012ig} leads to significantly larger predictions for 
the cross sections.  The corresponding plots for the $s$-channel and
associated $\Wboson\topquark$ production are given in Figs.\ 
\ref{fig:app-alphasdep-s}, \ref{fig:app-alphasdep-Wt} in
App.~\ref{sec:app-addfigs}.\\

\begin{figure}[p!]
  \centering
  \subfloat[$t$-channel production]{%
    \includegraphics[width=1.\textwidth]{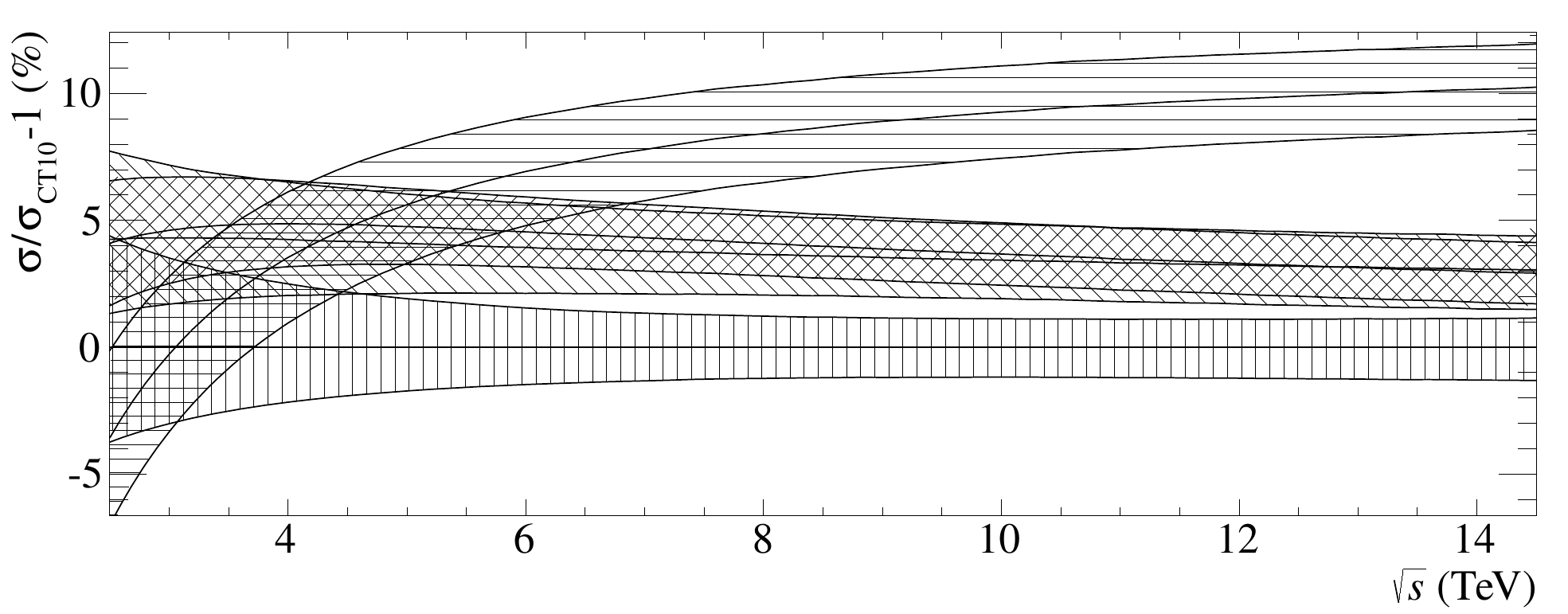}
    \label{fig:pdfuncert-a}
  } \\
  \subfloat[$s$-channel production]{%
    \includegraphics[width=1.\textwidth]{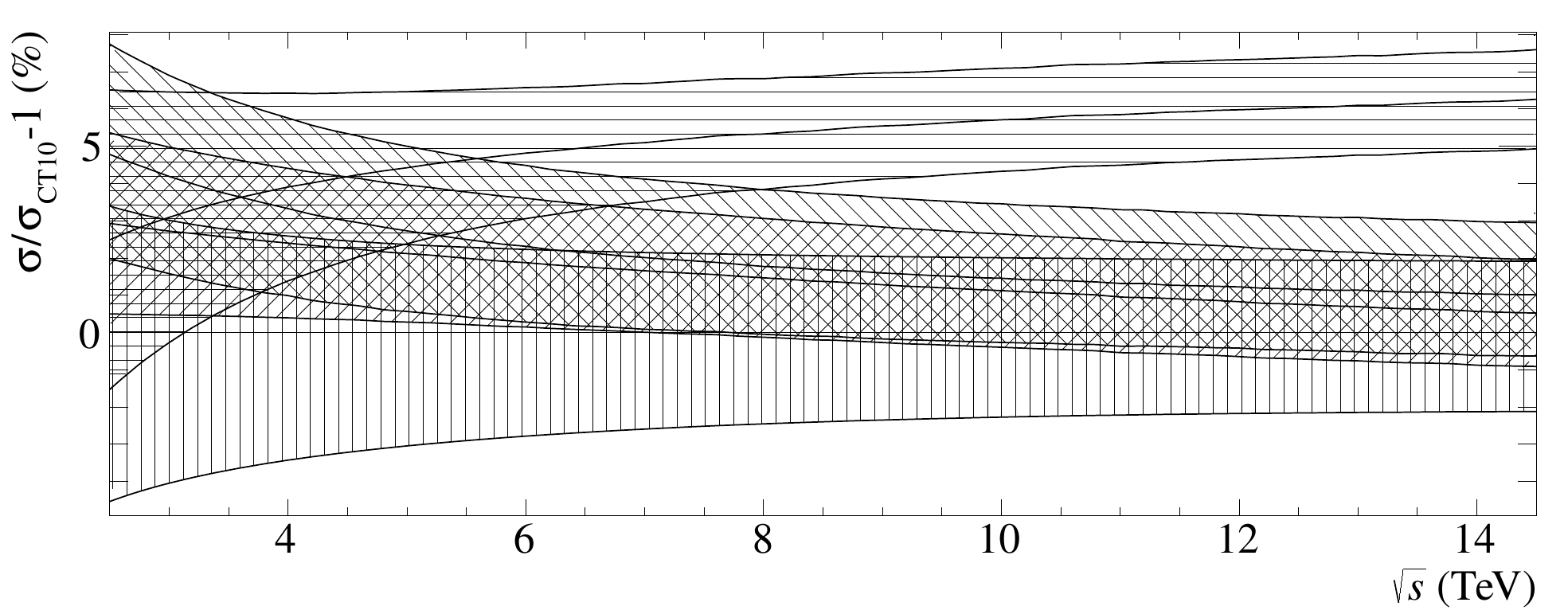}
    \label{fig:pdfuncert-b}
  } \\
  \subfloat[Associated $\Wboson\topquark$ production]{%
    \includegraphics[width=1.\textwidth]{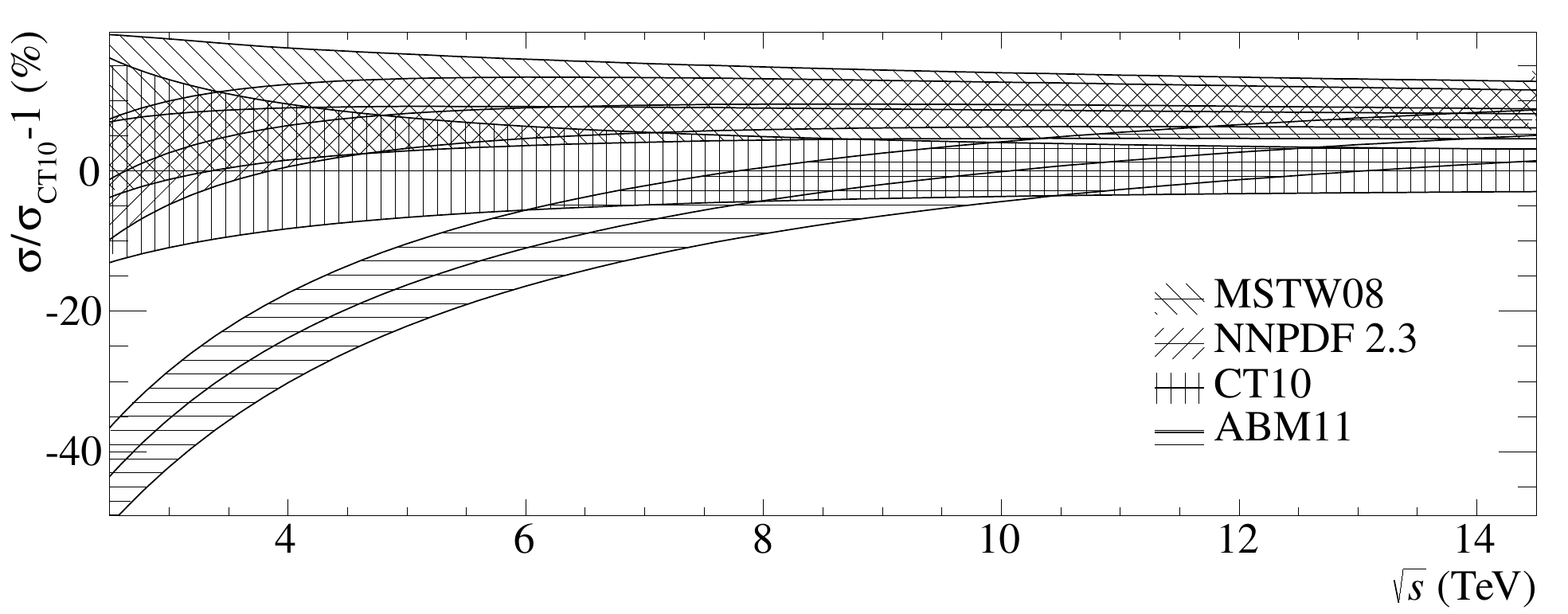}
    \label{fig:pdfuncert-c}
  }
  \caption{PDF uncertainties for the predicted cross sections for
    single top-quark production in proton-proton scattering as
    function of the
    centre-of-mass energy,\  $\sqrt{s}$. The uncertainty bands are
    calculated according to\ \cite{Alekhin:2012ig} for ABM11,
    \cite{Lai:2010vv} for CT10, \cite{Martin:2009iq} for MSTW08
    and\ \cite{Ball:2010de} for NNPDF23 (all PDFs in NLO accuracy). 
    All uncertainties are given
    with respect to CT10.}
  \label{fig:pdfuncert}
\end{figure}

\noindent\textit{PDF uncertainties} \\
Examples for the uncertainty related to our imperfect knowledge of the
parton distribution functions are shown \Fig{fig:pdfuncert} for the
three different channels contributing to single top-quark production.
For the PDF sets we used ABM11 \cite{Alekhin:2012ig}, 
CT10 \cite{Lai:2010vv}, MSTW08 \cite{Martin:2009iq} and NNPDF23 
\cite{Ball:2012cx}.
The uncertainty
bands include the uncertainties from the PDF fits as well as the
$\alphaS$ dependence. The results are normalised to the cross sections
calculated using CT10. The uncertainties for the individual PDF sets
are typically at the level of $\pm 2.5$ per cent at very low energies and
$\pm 1$ per cent at high energies in case of the $s$ and $t$-channels. In
case of associated $\Wt$ production the uncertainties are significantly larger.
Comparing different PDF sets, the CT10 predictions are slightly
below the predictions obtained using MSTW08 and NNPDF23. ABM11 on the
other hand leads to significantly larger cross sections at high
energies for the $t$-channel, while for the $\Wt$ production the cross section
is suppressed at low energies compared to other PDFs.
Within the uncertainty bands the results are, however, marginally consistent. 

Table\ \ref{tab:uncertainties} provides a breakdown of the
uncertainties for several PDF sets. The contributions from the
different sources of uncertainties are listed here for the case of
$t$-channel production in $\pp$ collisions at $\unit[14]{\TeV}$. \\

\renewcommand{\arraystretch}{1.2}
\begin{table}[b]
  \centering
  \begin{tabular}{@{}lrrrrr@{}}
        \toprule
        PDF Set & Cross section & PDF uncert. & $\alphaS$ uncert.
        & PDF+$\alphaS$ uncert. & Scale uncert. \\
        \midrule
        ABM11        & \unit[267.0]{pb}
                     & $\unit[\pm1.3]{\%}$
                     & $\unit[\pm0.8]{\%}$
                     & $\unit[\pm1.6]{\%}$
                     & $^{\unit[+2.8]{\%}}_{\unit[-1.6]{\%}}$ \\
        CT10         & \unit[242.4]{pb}
                     & $^{\unit[+0.9]{\%}}_{\unit[-1.1]{\%}}$
                     & $^{\unit[+0.7]{\%}}_{\unit[-0.8]{\%}}$
                     & $^{\unit[+1.1]{\%}}_{\unit[-1.3]{\%}}$
                     & $^{\unit[+3.0]{\%}}_{\unit[-1.7]{\%}}$ \\
        HERAPDF\,1.0 & \unit[247.3]{pb}
                     & $^{\unit[+5.0]{\%}}_{\unit[-1.3]{\%}}$
                     & $\unit[\pm0.9]{\%}$
                     & $^{\unit[+5.0]{\%}}_{\unit[-1.6]{\%}}$
                     & $^{\unit[+3.3]{\%}}_{\unit[-1.7]{\%}}$ \\
        MSTW 2008    & \unit[249.8]{pb}
                     & $\unit[\pm0.6]{\%}$
                     & $^{\unit[+1.2]{\%}}_{\unit[-1.4]{\%}}$
                     & $^{\unit[+1.3]{\%}}_{\unit[-1.5]{\%}}$
                     & $^{\unit[+3.0]{\%}}_{\unit[-1.6]{\%}}$ \\
        NNPDF\,2.3   & \unit[249.6]{pb} 
                     & $\unit[\pm0.5]{\%}$
                     & $\unit[\pm1.1]{\%}$
                     & $\unit[\pm1.2]{\%}$
                     & $^{\unit[+3.2]{\%}}_{\unit[-1.8]{\%}}$ \\
        \bottomrule
  \end{tabular}
  \caption{Sources of uncertainties for single top-quark production
                     cross sections at NLO for the
    $t$-channel in $\pp$ collisions with
    $\sqrt{s}\!=\!\unit[14]{\TeV}$ for various NLO PDF sets including
    the HERAPDF\
    \cite{Aaron:2009aa}.}
  \label{tab:uncertainties}
\end{table}
\renewcommand{\arraystretch}{1.0}

\section{Summary}
\label{sec:summary}

The extension of the \Hathor cross section calculator for the
production of single top-quarks has been presented. The new part
comprises a fast computation of all three production channels, the
$t$-channel, the $s$-channel and the associated $\Wboson\topquark$ production,
in LO and NLO of QCD. Despite the small amount of computing time
needed, the results agree within less than 1\,\textperthousand{} with
those from other programs. Moreover, the complete scale dependence for
$\muR$ and $\muF$ of the cross sections is implemented for all orders
up to NNLO and the corresponding equations are collected in this
paper.

Besides the scales, the \Hathor program allows to vary several model
parameters and the PDFs. This can be used to establish thorough
systematic studies of the total cross section in a fast and efficient
manner. Several examples were presented here, mostly for proton-proton
scattering at the \LHC. Among these are cross section calculations in
different orders and the dependence on  renormalisation and
factorisation scales. The dependence on the mass of the top quark was
studied in some detail and estimates of the top-quark mass from
single top-quark production using the latest cross section measurements were
given. Furthermore, the impact of the strong coupling constant,
$\alphaS$, and the parton distribution functions on the cross section
uncertainty has been discussed. 

\section*{Acknowledgments}

We would like to thank our colleagues from Wuppertal University for
their constructive feedback on preliminary versions of the \Hathor
program. Furthermore, we thank Sven-Olaf Moch for a careful reading of
the manuscript. This work is partially supported by the Helmholtz
Alliance ``Physics at the Terascale'' HA-101 and by the German Federal
Ministry for Education and Research (05H12KHE). P.~Rieck is supported
by the ``Studienstiftung des deutschen Volkes''.

\appendix
\section{Scaling functions in NNLO}
\label{sec:app-scalingNNLO}
Using 
\begin{equation}
  P_{{\quark_i}{\quark_j}} = \delta_{ij} P^V_{\quark\quark} 
  + P^S_{\quark\quark},
\end{equation}
\begin{equation}
  P_{{\quark_i}{\antiquark_j}} = \delta_{ij}
  P^V_{\quark\antiquark} 
  + P^S_{\quark\antiquark},
\end{equation}
the scale dependent terms for
$s$-channel read ($i\in\{20,21,22\}$):
\allowdisplaybreaks[4] 
\begin{align*}
\ff{i}{\quark\quark'}
&=
\FF{i}{\upquark\antidown}\, \absV{\quark\quark'}^2\,
    \Vsum{\topquark}
    + \FF{i}{\upquark\charmquark}\, \left(\Vsum{\quark}+\Vsum{\quark'}\right)\,
    \Vsum{\topquark}
 & \quark &\in\begin{aligned}[t]\{&
\upquark,\charmquark\}
\end{aligned},\quark' \in \{\antib,\antistrange,\antidown\},
\\
\ff{i}{\quark\quark'}
&=
\FF{i}{\upquark\charmquark}\,\Vsum{\quark}\,\Vsum{\topquark}
    + \FF{i}{\upquark\downquark}\,\absV{\quark\quark'}^2\,
    \Vsum{\topquark}
 & \quark\quark' &\in\begin{aligned}[t]\{&
 \charmquark\bquark, \charmquark\downquark, \charmquark\strangequark,
 \upquark\bquark, \upquark\downquark, \upquark\strangequark, \antib\antiup,
 \antib\anticharm, \antistrange\antiup, \antistrange\anticharm,
 \antidown\antiup, \antidown\anticharm 
 \}
\end{aligned}
\\
\ff{i}{\quark\quark'}
&=
\FF{i}{\upquark\charmquark}\,\left(\Vsum{\quark}+ \Vsum{\quark'}\right)\,\Vsum{\topquark}
 & \quark\quark' &\in\begin{aligned}[t]\{&
\upquark\charmquark, \upquark\upquark, \charmquark\charmquark,
\antidown\antidown, \antistrange\antistrange, \antib\antib,
\antidown\antistrange, \antidown\antib, \antistrange\antib
\}
\end{aligned}
\\
\ff{i}{\quark\quark'}
&=
\FF{i}{\upquark\charmquark}\,\Vsum{\quark}\,\Vsum{\topquark}
 & \quark\quark' &\in\begin{aligned}[t]\{&
\upquark\antiup, \upquark\anticharm, \charmquark\antiup, \charmquark\anticharm,
 \antib\bquark, \antib\downquark, \antib\strangequark, \antistrange\bquark, \antistrange\downquark, \antistrange\strangequark, \antidown\bquark, \antidown\downquark, 
\antidown\strangequark
\}
\end{aligned}
\\
\ff{i}{\gluon\quark}
&=
\FF{i}{\gluon\upquark}\, \Vsum{\quark}\, \Vsum{\topquark}
 & \quark &\in\begin{aligned}[t]\{&
\antib, \antidown, \antistrange, \charmquark, \upquark\}
\end{aligned}
\\
\ff{i}{\gluon\gluon}
&=
\FF{i}{\gluon\gluon}\,
    \left(\Vsum{\upquark}+\Vsum{\charmquark}\right)\,
    \Vsum{\topquark} &&& 
\end{align*}
with
  \allowdisplaybreaks[4]
  \begin{alignat*}{2}
    \FF{21}{\upquark\antidown} &=& 
    -2&\,\PqqV{1} \,\otimes\,\FF{0}{\upquark\antidown}
    +\left(\beta_0 - 2\Pqq{0}\right)\,\otimes\,\FF{10}{\upquark\antidown}\,, \\
    \FF{22}{\upquark\antidown} &=& &
    \left(-\beta_0\Pqq{0} + 2\Pqq{0}\,\otimes\,\Pqq{0}\right)
    \,\otimes\,\FF{0}{\upquark\antidown}\,, \hfill \\
    \FF{21}{\gluon\upquark} &=&
    -&\Pqg{1}\,\otimes\,\FF{0}{\upquark\antidown}
    -\Pqg{0}\,\otimes\,\FF{10}{\upquark\antidown}
    +\left(\beta_0-\Pqq{0}-\Pgg{0}\right)\,\otimes\,\FF{10}{\gluon\upquark}\,, \\
    \FF{22}{\gluon\upquark} &=&
    \dfrac{1}{2}&\left(3 \Pqq{0}\,\otimes\,\Pqg{0} - \beta_0\Pqg{0} +
                      \Pqg{0}\,\otimes\,\Pgg{0}
                \right)\,\otimes\,\FF{0}{\upquark\antidown} \,, \\
    \FF{21}{\gluon\gluon} &=&
    -4&\, \Pqg{0}\,\otimes\,\FF{10}{\gluon\upquark} \,, \\
    \FF{22}{\gluon\gluon} &=&
    2&\, \Pqg{0}\,\otimes\,\Pqg{0}\,\otimes\,\FF{0}{\upquark\antidown} \,, \\
    \FF{21}{\upquark\charmquark} &=& 
    -&\PqqS{1}\,\otimes\,\FF{0}{\upquark\antidown}
    -\Pgq{0}\,\otimes\,\FF{10}{\gluon\upquark} \,, \\
    \FF{22}{\upquark\charmquark} &=& 
    \dfrac{1}{2}&\, \Pqg{0}\,\otimes\,\Pgq{0}\,\otimes\,\FF{0}{\upquark\antidown}
    \,, \\
    \FF{21}{\upquark\downquark} &=& 
    -&\PqqbV{1}\,\otimes\,\FF{0}{\upquark\antidown}\,, \\
    \FF{22}{\upquark\downquark} &=& & 0 \,.
  \end{alignat*}

\noindent $t$-channel:
    \allowdisplaybreaks[4] 
    \begin{align*}
\ff{i}{\quark\quark'}
&=\FF{i}{\bquark\upquark}\, \Vsum{\quark}\absV{\topquark\quark'}^2
+\FF{i}{\upquark\charmquark}\, \Vsum{\quark}\, \Vsum{\topquark} 
+\FF{i}{\bquark\downquark}\, (\Vsum{\upquark}+\Vsum{\charmquark})\absV{\topquark\quark'}^2\, 
& \quark &\in\begin{aligned}[t]\{&
\charmquark,\upquark\},\quark'\in\{\downquark,\strangequark,\bquark\}
\end{aligned}
\\
\ff{i}{\quark\quark'}
&=\FF{i}{\bquark\antidown}\, \Vsum{\quark}\absV{\topquark\quark'}^2
+\FF{i}{\antidown\antiup}\, \Vsum{\quark}\, \Vsum{\topquark} 
+\FF{i}{\bquark\downquark}\, (\Vsum{\upquark}+\Vsum{\charmquark})\absV{\topquark\quark'}^2\, 
& \quark &\in\begin{aligned}[t]\{&
\antidown,\antistrange,\antib\},\quark'\in\{\downquark,\strangequark,\bquark\}
\end{aligned}
\\
\ff{i}{\quark\quark'}
&=\FF{i}{\bquark\antiup}\, \Vsum{\quark}\absV{\topquark\quark'}^2
+\FF{i}{\bquark\downquark}\, (\Vsum{\upquark}+\Vsum{\charmquark})\absV{\topquark\quark'}^2
\, 
& \quark &\in\begin{aligned}[t]\{&
\anticharm,\antiup\},\quark'\in\{\downquark,\strangequark,\bquark\}
\end{aligned}
\\
\ff{i}{\quark\quark'}
&=\FF{i}{\bquark\antiup}\, \Vsum{\quark}\absV{\topquark\quark'}^2
+\FF{i}{\upquark\charmquark}\, \Vsum{\quark}\, \Vsum{\topquark} 
+\FF{i}{\antidown\antiup}\, \Vsum{\quark'}\Vsum{\topquark}\, 
& \quark &\in\begin{aligned}[t]\{&
\charmquark,\upquark\},\quark'\in\{\antidown,\antistrange,\antib\}
\end{aligned}
\\
\ff{i}{\quark\quark'}
&=\FF{i}{\antidown\antib}\,(
\Vsum{\quark}\absV{\topquark\quark'}^2+\Vsum{\quark'}\absV{\topquark\quark}^2)
+\FF{i}{\bquark\downquark}\, (\Vsum{\upquark}+\Vsum{\charmquark})
(\absV{\topquark\quark'}^2+\absV{\topquark\quark}^2)
\, 
& \quark &\in\begin{aligned}[t]\{&
\downquark,\strangequark,\bquark\},\quark'\in\{\downquark,\strangequark,\bquark\}
\end{aligned}
\\
\ff{i}{\quark\quark'}
&=\FF{i}{\antidown\antiup}\, (\Vsum{\quark}+\Vsum{\quark'})\, \Vsum{\topquark} 
+\FF{i}{\antidown\antib}\, (\Vsum{\quark}\absV{\topquark\quark'}^2
+\Vsum{\quark'}\absV{\topquark\quark}^2 )\, 
& \quark &\in\begin{aligned}[t]\{&
\antidown,\antistrange,\antib\},\quark'\in\{\antidown,\antistrange,\antib\}
\end{aligned}
\\
\ff{i}{\quark\quark'}
&=\FF{i}{\antidown\antib}\, \Vsum{\quark}\, \Vsum{\topquark} 
\, 
& \quark &\in\begin{aligned}[t]\{&
\antidown,\antistrange,\antib\},\quark'\in\{\antiup,\anticharm\}
\end{aligned}
\\
\ff{i}{\quark\quark'}
&=\FF{i}{\upquark\charmquark}\, \Vsum{\quark}\, \Vsum{\topquark} 
\, 
& \quark &\in\begin{aligned}[t]\{&
\upquark,\charmquark\},\quark'\in\{\antiup,\anticharm\}
\end{aligned}
\\
\ff{i}{\quark\quark'}
&=\FF{i}{\upquark\charmquark}\, (\Vsum{\quark}+\Vsum{\quark'}   )\, \Vsum{\topquark} 
\, 
& \quark &\in\begin{aligned}[t]\{&
\upquark,\charmquark\},\quark'\in\{\upquark,\charmquark\}
\end{aligned}
\\
\ff{i}{\gluon\quark}
&=
\FF{i}{\gluon\bquark}\, (\Vsum{\upquark} + \Vsum{\charmquark})\, \absV{\topquark\quark}^2
 & \quark &\in\begin{aligned}[t]\{&
\bquark, \downquark, \strangequark\}
\end{aligned}
\\
\ff{i}{\gluon\quark}
&=
\FF{i}{\gluon\antib}\, \Vsum{\antiquark}\, \Vsum{\topquark}
 & \quark &\in\begin{aligned}[t]\{&
\antib, \antidown, \antistrange\}
\end{aligned}
\\
\ff{i}{\gluon\quark}
&=
\FF{i}{\gluon\upquark}\, \Vsum{\quark}\, \Vsum{\topquark}
 & \quark &\in\begin{aligned}[t]\{&
\charmquark, \upquark\}
\end{aligned}
\\
\ff{i}{\gluon\gluon}
&=
\FF{i}{\gluon\gluon}\, (\Vsum{\upquark} + \Vsum{\charmquark})\, \Vsum{\topquark} &&&
  \end{align*}
with
  \allowdisplaybreaks[4] 
  \begin{alignat*}{2}
   \FF{21}{\bquark\upquark} &=&
   -2&\,\PqqV{1} \,\otimes\,\FF{0}{\bquark\upquark}\, 
   +  \,\left(\beta_0 - 2\,\Pqq{0}\right)\,\otimes\,\FF{10}{\bquark\upquark}\,, \\
   \FF{22}{\bquark\upquark} &=&
   &  \left(-\beta_0\,\Pqq{0}\,+\,2\,\Pqq{0}\,\otimes\,\Pqq{0}\right)\,\otimes\,
      \FF{0}{\bquark\upquark}\,, \\
   \FF{21}{\bquark\antidown} &=&
   -2&\,\PqqV{1} \,\otimes\,\FF{0}{\bquark\antidown}\, 
   +  \,\left(\beta_0 - 2\,\Pqq{0}\right)\,\otimes\,
   \FF{10}{\bquark\antidown}\,, \\
   \FF{22}{\bquark\antidown} &=&
   &  \left(-\beta_0\,\Pqq{0}\,+\,2\,\Pqq{0}\,\otimes\,\Pqq{0}\right)
   \,\otimes\,\FF{0}{\bquark\antidown}\,, \\
   \FF{21}{\gluon\upquark} &=&
   - &\Pqg{1}\,\otimes\,\FF{0}{\bquark\upquark}\, 
   -\,\Pqg{0}\,\otimes\,\FF{10}{\bquark\upquark}\,
   +\,\left(\beta_0\,-\,\Pqq{0}\,-\,\Pgg{0}\right)\,\otimes\,
   \FF{10}{\gluon\upquark}\\
   \FF{22}{\gluon\upquark} &=&
   \dfrac{1}{2} & \left(3\,\Pqq{0}\,\otimes\,\Pqg{0}\,-\,\beta_0\,\Pqg{0}\,
   +\,\Pqg{0}\,\otimes\,\Pgg{0}\right)\,\otimes\,\FF{0}{\bquark\upquark}\,, \\
   \FF{21}{\gluon\antidown} &=&
   - &\Pqg{1}\,\otimes\,\FF{0}{\bquark\antidown}\, 
   -\,\Pqg{0}\,\otimes\,\FF{10}{\bquark\antidown}\,
   +\,(\beta_0\,-\,\Pqq{0}\,-\,\Pgg{0})\,\otimes\,\FF{10}{\gluon\antiup}\\
   \FF{22}{\gluon\antidown} &=&
   \dfrac{1}{2} & (3\,\Pqq{0}\,\otimes\,\Pqg{0}\,-\,\beta_0\,\Pqg{0}\,
   +\,\Pqg{0}\,\otimes\,\Pgg{0})
   \,\otimes\,\FF{0}{\bquark\antidown}\,, \\
   \FF{21}{\gluon\bquark} &=&
   - &\Pqg{1}\,\otimes\,\FF{0}{\bquark\upquark}\, 
   -\,\Pqg{0}\,\otimes\,\FF{10}{\bquark\upquark}\,
   -  \Pqg{1}\,\otimes\,\FF{0}{\bquark\antidown}\, 
   -\,\Pqg{0}\,\otimes\,\FF{10}{\bquark\antidown}\, \nonumber \\
& & + & \left(\beta_0\,-\,\Pqq{0}\,-\,\Pgg{0}\right)\,\otimes\,
   \FF{10}{\gluon\bquark}\\
   \FF{22}{\gluon\bquark} &=&
   & \FF{22}{\gluon\upquark}\,+\,\FF{22}{\gluon\antidown}\,, \\
   \FF{21}{\gluon\gluon} &=&
   -2 &\,\Pqg{0}\,\otimes\,\FF{10}{\gluon\upquark}\, 
   -  2\,\Pqg{0}\,\otimes\,\FF{10}{\gluon\bquark}\, 
   -  2\,\Pqg{0}\,\otimes\,\FF{10}{\gluon\antidown}\,, \\
   \FF{22}{\gluon\gluon} &=&
   2 &\, \Pqg{0}\,\otimes\,\Pqg{0}\,\otimes\,\FF{0}{\bquark\upquark}\, + 
   2 \,\Pqg{0}\,\otimes\,\Pqg{0}\,\otimes\,\FF{0}{\bquark\antidown} \,, \\
   \FF{21}{\upquark\charmquark} &=&
   - &\,\PqqS{1}\,\otimes\,\FF{0}{\bquark\upquark}\,
   -\,\Pgq{0}\,\otimes\,\FF{10}{\gluon\upquark} \,, \\
   \FF{22}{\upquark\charmquark} &=&
   \frac{1}{2} &\,\Pqg{0}\,\otimes\,\Pgq{0}\,\otimes\,\FF{0}{\bquark\upquark}\,,\\
   \FF{21}{\antidown\antiup} &=&
   - &\,\PqqS{1}\,\otimes\,\FF{0}{\bquark\antidown}\,
   -\,\Pgq{0}\,\otimes\,\FF{10}{\gluon\antidown} \,, \\
   \FF{22}{\antidown\antiup} &=&
   \frac{1}{2} &\,\Pqg{0}\,\otimes\,\Pgq{0}\,\otimes\,\FF{0}{\bquark\antidown}
    \,, \\
   \FF{21}{\bquark\downquark} &=&
    - &\,\PqqS{1}\,\otimes\,\FF{0}{\bquark\upquark}\, 
   -   \,\PqqS{1}\,\otimes\,\FF{0}{\bquark\antidown}\, 
   -   \,\Pgq{0}\,\otimes\,\FF{10}{\gluon\bquark}  \,, \\
   \FF{22}{\bquark\downquark} &=&
   \dfrac{1}{2} & \left(
   \Pqg{0}\,\otimes\,\Pgq{0}\,\otimes\,\FF{0}{\bquark\upquark} \,
   + \Pqg{0}\,\otimes\,\Pgq{0}\,\otimes\,\FF{0}{\bquark\antidown}
   \right) \,, \\
   \FF{21}{\antidown\antib} &=&
   - &\,\PqqbV{1}\,\otimes\,\FF{0}{\bquark\antidown}  \,, \\
   \FF{21}{\bquark\antiup} &=&
   - &\,\PqqbV{1}\,\otimes\,\FF{0}{\bquark\upquark}  \,, \\
   \FF{22}{\antidown\antib} &=&
     & \FF{22}{\bquark\antiup} = 0 \,. 
  \end{alignat*}

\noindent Associated $\Wboson\topquark$ production, $i \in \{20,21,22\}$,$\kappa=0$
for all partonic channels. If $\bquark\bquark$ and $\antib\bquark$
are not included, $f_{gb}^{(i)}$ and $f_{g\antib}^{(i)}$ change ($\kappa=1$):
    \allowdisplaybreaks[4] 
    \begin{align*}
\ff{i}{\gluon\quark}
&=\FF{i}{\gluon\bquark}\, \absV{\topquark\quark}^2
+\FF{i}{\gluon\upquark}\, \Vsum{\topquark}&
  \quark &\in\begin{aligned}[t]\{&
\bquark, \downquark, \strangequark\}
\end{aligned}
\\
\ff{i}{\gluon\antiquark}
&=\FF{i}{\gluon\antib}\, \absV{\topquark\antiquark}^2
+\FF{i}{\gluon\upquark}\, \Vsum{\topquark}
 & \antiquark &\in\begin{aligned}[t]\{&
\antib, \antidown, \antistrange\}
\end{aligned}
\\
\ff{i}{\quark\quark}
&=
2\FF{i}{\bquark\upquark}\, \absV{\topquark\quark}^2
 & \quark &\in\begin{aligned}[t]\{&
\downquark,\strangequark,\bquark\},
\end{aligned}
\\
\ff{i}{\quark\quark'}
&=
\FF{i}{\bquark\upquark}\, \absV{\topquark\quark'}^2
& \quark &\in\begin{aligned}[t]\{&
  \antib,\anticharm,\antistrange,\antiup,\antidown,\downquark,\upquark,
  \strangequark,\charmquark,\bquark\},
  \quark' \in \{\downquark,\strangequark,\bquark\},\quark \ne \quark'
\end{aligned}
\\
\ff{i}{\gluon\quark}
&=
\FF{i}{\gluon\upquark}\, \Vsum{\topquark}
 & \quark &\in\begin{aligned}[t]\{&
   \anticharm, \antiup, \charmquark, \upquark\}
 \end{aligned}
 \\
\ff{i}{\gluon\gluon}
&=
\FF{i}{\gluon\gluon}\, \Vsum{\topquark} &&&
    \end{align*}
with
  \allowdisplaybreaks[4] 
  \begin{alignat*}{2}
   \FF{21}{\gluon\bquark} &=& 
   -(2 & \,n_{\text{f}}-3\kappa) \Pqg{0}\,\otimes\,\FF{10}{\bquark\upquark}\,
   +\,\left(\beta_1\,-\,\Pgg{1}\,-\,\PqqV{1}\right)
   \,\otimes\,\FF{0}{\gluon\bquark} \nonumber \\
& & & +\,\left(2\beta_0\,-\,\Pqq{0}\,-\,\Pgg{0}\right)
   \,\otimes\,\FF{10}{\gluon\bquark} \,, \\
   \FF{22}{\gluon\bquark} &=& 
    & \left(\beta_0^2\,-\,\dfrac{3}{2}\beta_0\Pqq{0} \,
    +\,\dfrac{1}{2}\Pqq{0}\,\otimes\,\Pqq{0}\,
    +\,\Pqq{0}\,\otimes\,\Pgg{0}\,+\frac{1}{2}(2\,n_{\text{f}}-3\kappa)\,\Pqg{0}\,\otimes\,\Pgq{0}
    \right. \nonumber \\
& & & \left. -\,\dfrac{3}{2}\beta_0\Pgg{0}\,
    +\,\dfrac{1}{2}\Pgg{0}\,\otimes\,\Pgg{0}
    \right)\,\otimes\,\FF{0}{\gluon\bquark} \,, \\
   \FF{21}{\gluon\gluon} &=&
   -2 &\, \Pqg{1}\,\otimes\,\FF{0}{\gluon\bquark}\,
   - 2 \, \Pqg{0}\,\otimes\,\FF{10}{\gluon\bquark}\,
   +2\left(\beta_0\,-\,\Pgg{0}\right)\,\otimes\,\FF{10}{\gluon\gluon}\,, \\
   \FF{22}{\gluon\gluon} &=&
     &\, \left(\Pqq{0}\,\otimes\,\Pqg{0}\,-\,3\beta_0\Pqg{0}\,+\,
     3\Pqg{0}\,\otimes\,\Pgg{0}\right)\,\otimes\,\FF{0}{\gluon\bquark} \,, \\
   \FF{21}{\bquark\upquark} &=&
   & \left(2\beta_0\,-\,2\Pqq{0}\right)\,\otimes\,\FF{10}{\bquark\upquark}\,
   -\,\Pgq{1}\,\otimes\,\FF{0}{\gluon\bquark}
   -\,\Pgq{0}\,\otimes\,\FF{10}{\gluon\bquark} \,, \\
   \FF{22}{\bquark\upquark} &=&
   \dfrac{1}{2} & \left(3\,\Pqq{0}\,\otimes\,\Pgq{0}\,-\,3\,\beta_0\Pgq{0}\,
   +\,\Pgq{0}\,\otimes\,\Pgg{0}\right)\,\otimes\,\FF{0}{\gluon\bquark} \,, \\
   \FF{21}{\gluon\upquark} &=&
   -&\,\Pqg{0}\,\otimes\,\FF{10}{\bquark\upquark}\,
   - \,\PqqS{1}\,\otimes\,\FF{0}{\gluon\bquark}\,
   - \,\Pgq{0}\,\otimes\,\FF{10}{\gluon\gluon} \,, \\
   \FF{22}{\gluon\upquark} &=&
   \frac{3}{2} &\,\Pqg{0}\,\otimes\,\Pgq{0}\,\otimes\,\FF{0}{\gluon\bquark}\,,\\
   \FF{21}{\gluon\antib} &=&
   - &\,\PqqbV{1}\,\otimes\,\FF{0}{\gluon\bquark}\,
   + \kappa\, \Pqg{0}\,\otimes\,\FF{10}{\bquark\upquark} \,, \\
   \FF{22}{\gluon\antib} &=&
   -&\frac{\kappa}{2}\Pqg{0}\,\otimes\,\Pgq{0}\,\otimes\,\FF{0}{\gluon\bquark} \,.
 \end{alignat*}

\newpage

\section{Additional figures}
\label{sec:app-addfigs}

\begin{figure}[ht!]
  \centering
  \subfloat[$s$-channel production]{%
    \includegraphics[width=0.7\textwidth]{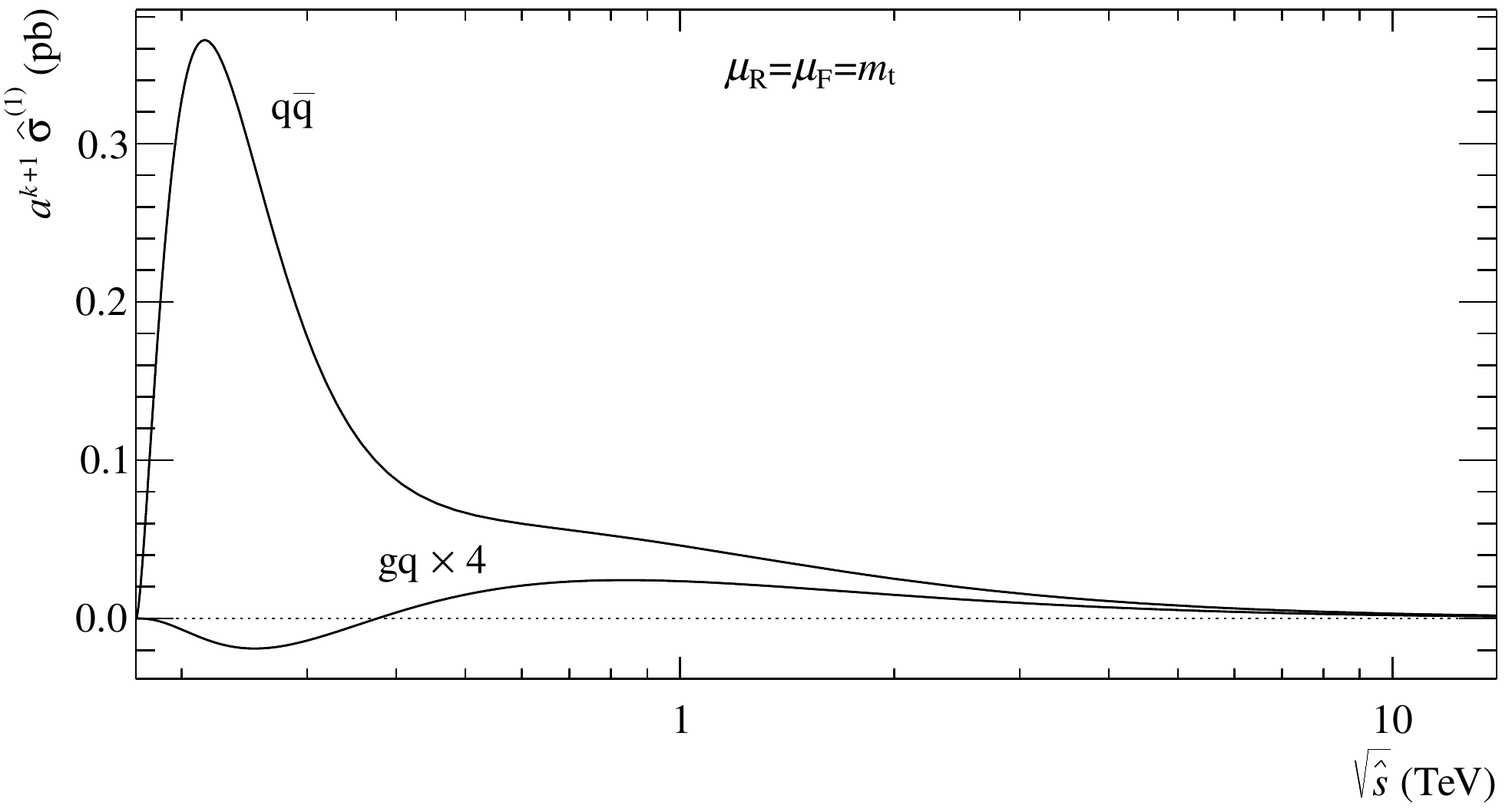}
    \label{fig:app-partxsdep-a}
  } \\
  \subfloat[Associated $\Wboson\topquark$ production]{%
    \includegraphics[width=0.7\textwidth]{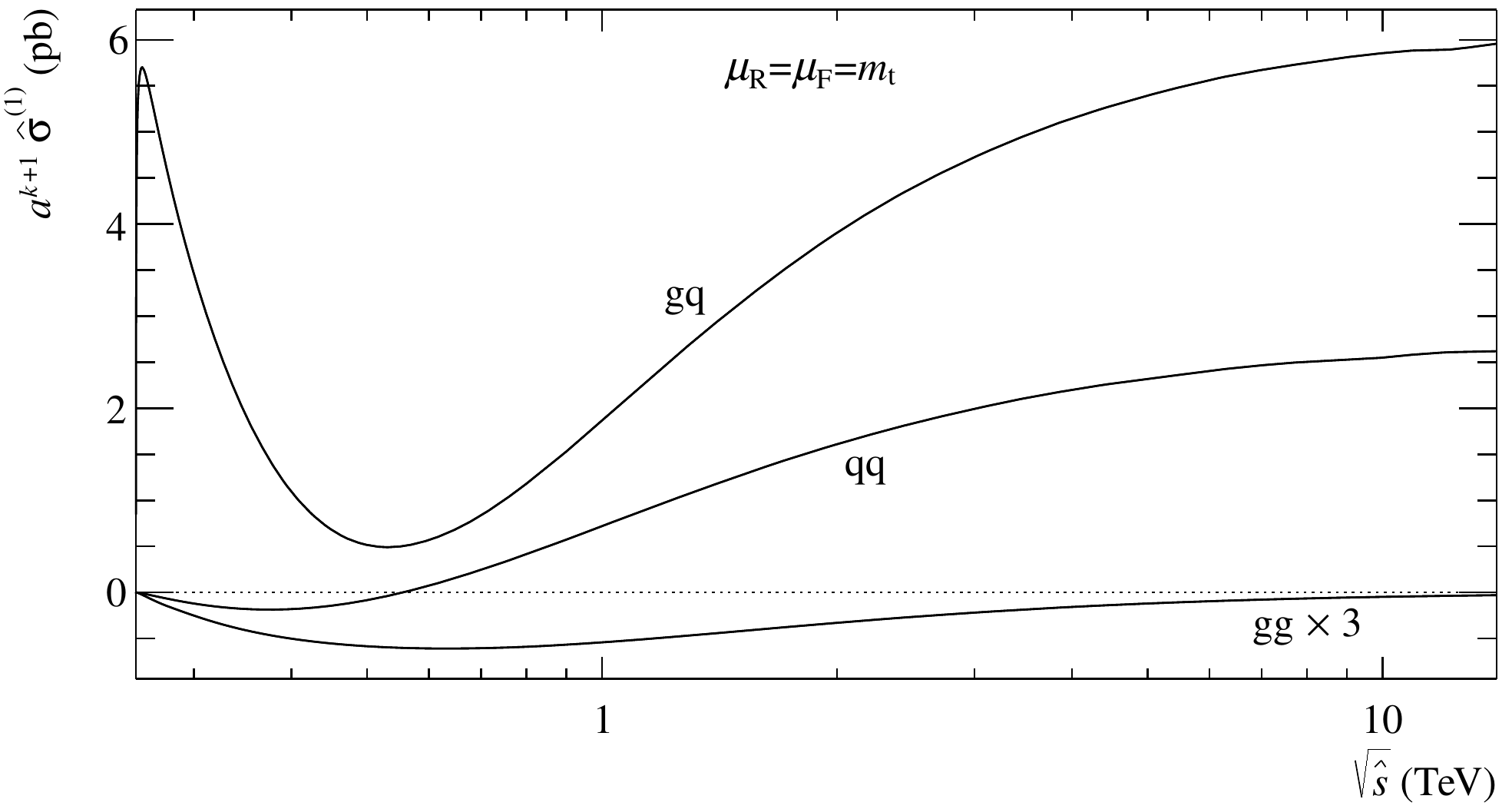}
    \label{fig:app-partxsdep-b}
  }
  \caption{NLO contributions to the partonic cross section for single
    top-quark production ($s$-channel, upper plot)
    ($\Wboson\topquark$ channel, lower plot) as a function of the
    partonic centre-of-mass energy for a top-quark mass of
    \unit[172.5]{\GeV}. The scales $\muR$ and $\muF$ are set to the
    mass of the top quark. Here, $\quark$ and $\antiquark$ indicate
    all applicable flavours for the given channel.}
  \label{fig:app-partonicxs}
\end{figure}

\begin{figure}[ht!]
  \centering
  \subfloat[$s$-channel production]{%
    \includegraphics[width=0.75\textwidth]{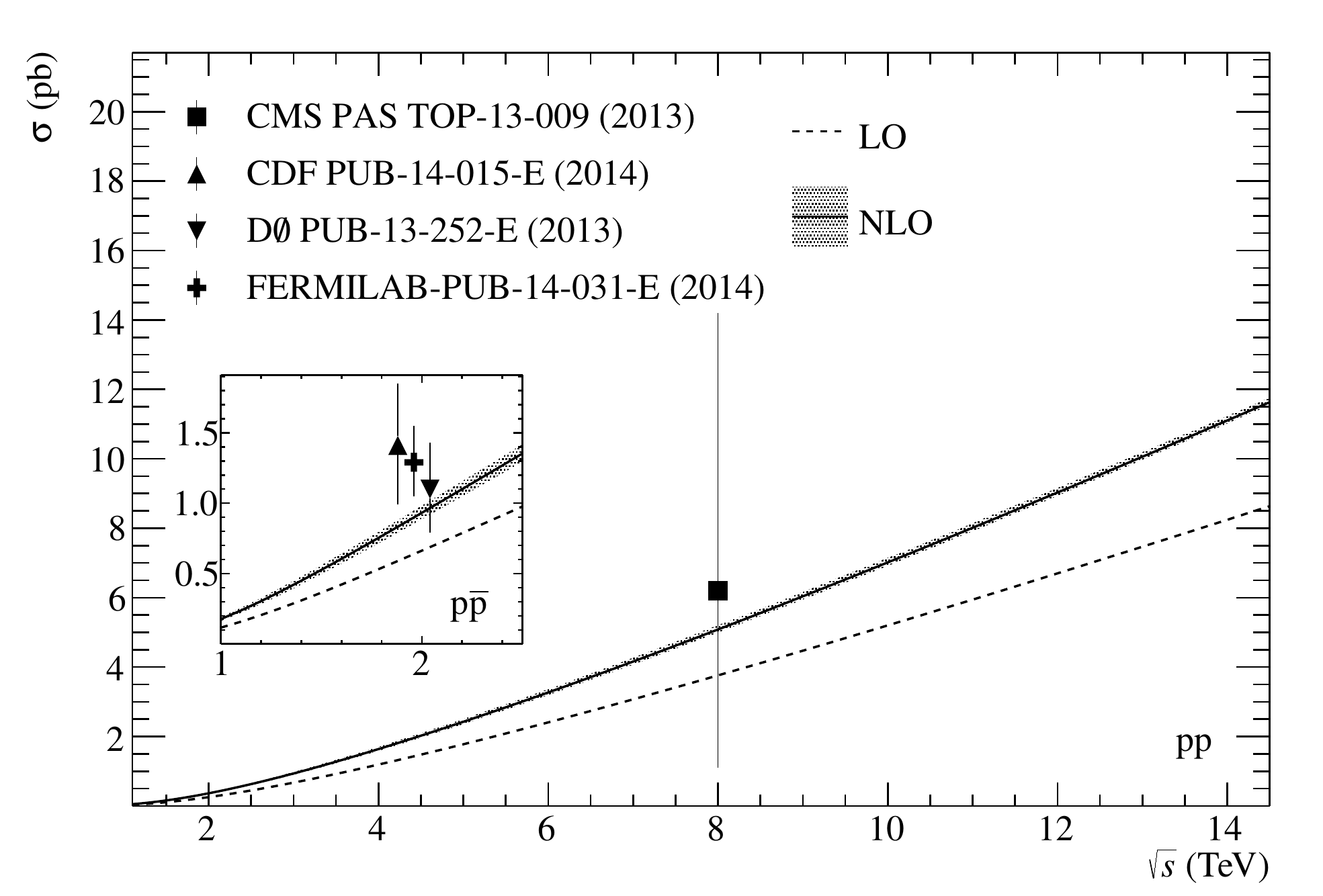}
    \label{fig:theouncert-s}
  } \\
  \subfloat[Associated $\Wboson\topquark$ production]{%
    \includegraphics[width=0.75\textwidth]{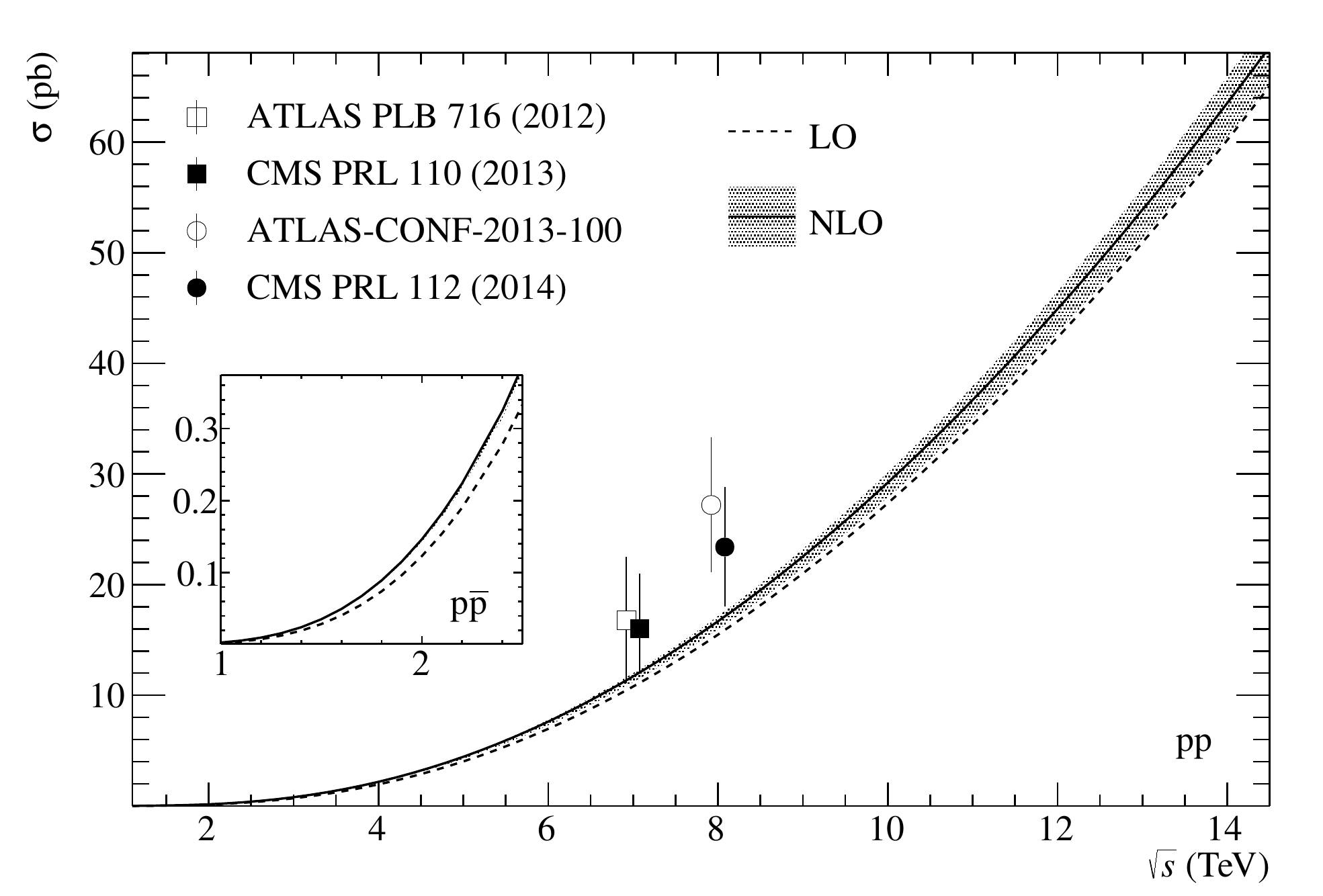}
    \label{fig:theouncert-Wt}
  }
  \caption{LO and NLO cross sections for single top-quark production calculated
    with the HATHOR program using CT10nlo as PDF set. The uncertainty
    band shown for NLO indicates the scale uncertainty. 
    The latest measurements from
    \CMS\ \cite{CMS-PAS-TOP-13-009} and the \Tevatron experiments\ 
    \cite{Aaltonen:2014qja,D0:2013,CDF:2014uma} for $s$-channel
    production are shown for comparison, as well as measurements from
    the \LHC for associated $\Wboson\topquark$
    production\ \cite{Aad:2012dj,PhysRevLett.110.022003,%
      ATLAS-CONF-2013-100,CMS-PAS-TOP-12-040}.}
  \label{fig:theouncert-sWt}
\end{figure}

\begin{figure}[ht!]
  \centering
  \subfloat[$s$-channel production]{%
    \includegraphics[width=0.7\textwidth]{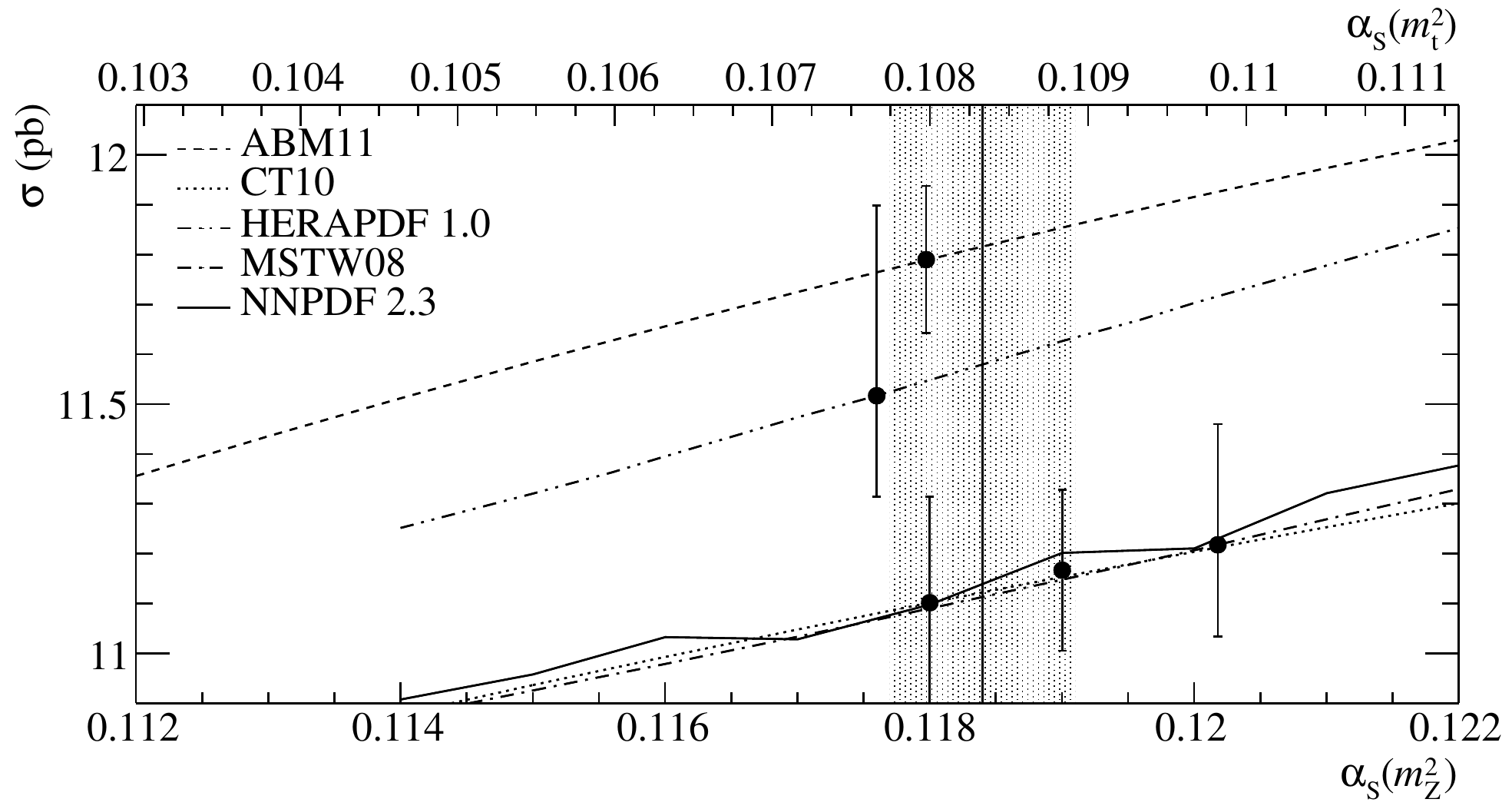}
    \label{fig:app-alphasdep-s}
  } \\
  \subfloat[Associated $\Wboson\topquark$ production]{%
    \includegraphics[width=0.7\textwidth]{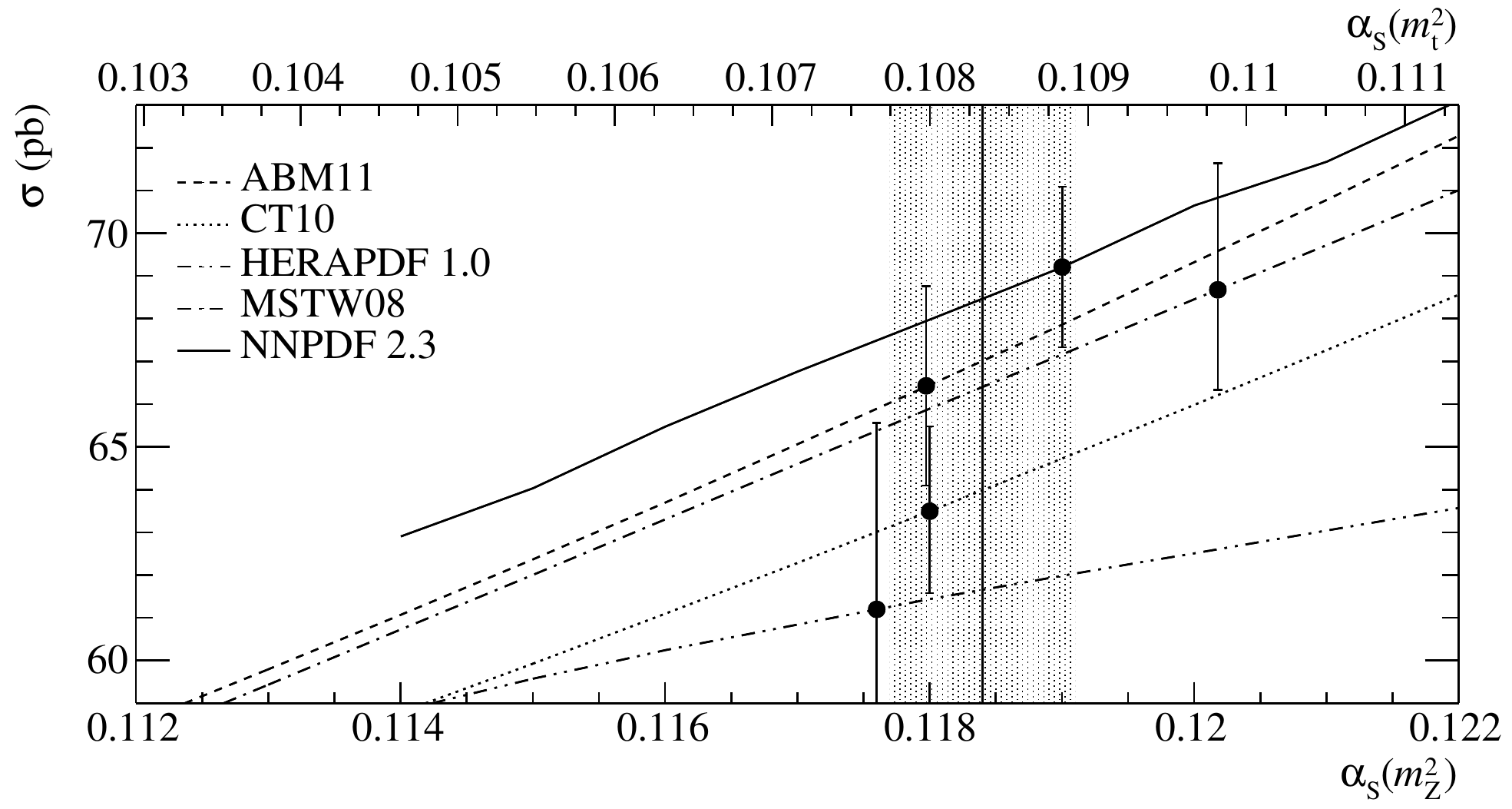}
    \label{fig:app-alphasdep-Wt}
  }
  \caption{Single top-quark cross sections for the $s$-channel and
    $\Wboson\topquark$ production in NLO for different values of
    $\alphaS$, computed for different NLO PDFs at
    $\sqrt{s}\!=\!\unit[14]{\TeV}$ in $\pp$ scattering. For each PDF
    set, the best fit value and the corresponding full PDF uncertainty
    is indicated by the black marker and the error bar. The $\alphaS$
    values are given with respect to the mass of the $\Zboson$ boson
    (lower abscissa) and the mass of the top quark (upper abscissa).
    The vertical line and the shaded box represent the latest world
    average value of $\alphaS$ and its uncertainty at the $\Zboson$ pole\
    \cite{Beringer:1900zz}.}
  \label{fig:app-alphasdep}
\end{figure}

\clearpage
\vspace*{-1.5cm} 
\enlargethispage{1.5cm}
\section{Parametrization of the top-quark mass
  dependence}
\label{sec:app-coeff}
\begin{sidewaystable}[H]
  \footnotesize
  \raggedleft
  \begin{tabular}{cccl*{12}{r@{.}l}}
    \toprule
    & & & & \multicolumn{6}{c}{ABM11}    & \multicolumn{6}{c}{CT\,10} 
    & \multicolumn{6}{c}{MSTW\,08} & \multicolumn{6}{c}{NNPDF\,2.3} \\
    \cmidrule(r){5-10}\cmidrule(r){11-16}
    \cmidrule(r){17-22}\cmidrule(r){23-28}
    & & & & \multicolumn{2}{c}{$\sigma(\bar{\mt})$} &
    \multicolumn{2}{c}{$A$} & \multicolumn{2}{c}{$B$} & 
    \multicolumn{2}{c}{$\sigma(\bar{\mt})$} &
    \multicolumn{2}{c}{$A$} & \multicolumn{2}{c}{$B$} & 
    \multicolumn{2}{c}{$\sigma(\bar{\mt})$} &
    \multicolumn{2}{c}{$A$} & \multicolumn{2}{c}{$B$} & 
    \multicolumn{2}{c}{$\sigma(\bar{\mt})$} &
    \multicolumn{2}{c}{$A$} & \multicolumn{2}{c}{$B$} \\
    \midrule
    \multirow{6}{*}{$\topquark$} & 
    \multirow{3}{*}{\begin{sideways}$\sqrt{s} = \unit[8]{\TeV}$\end{sideways}} & 
    \multirow{1}{*}{$t$} & 
    NLO & 59&2705 & -1&6541 & 1&569 & 53&5662 & -1&6082 & 1&507 & 55&1053 & -1&5995 & 1&492 & 55&2394 & -1&5932 & 1&479 \\
    \cmidrule{3-28}
    & & \multirow{1}{*}{$s$} & 
    NLO & 3&48131 & -3&8849 & 7&579 & 3&26836 & -3&8673 & 7&526 & 3&28661 & -3&8479 & 7&470 & 3&28760 & -3&8561 & 7&496 \\
    \cmidrule{3-28}
    & & \multirow{1}{*}{$\Wboson\topquark$} & 
    NLO & 7&99547 & -3&1930 & 5&61 & 8&34988 & -2&9994 & 4&591 & 9&9572 & -3&4 & 4&590 & 9&12887 & -3&118 & 4&637 \\
    \cmidrule{2-28}
    & \multirow{3}{*}{\begin{sideways}$\sqrt{s} = \unit[14]{\TeV}$\end{sideways}} & 
    \multirow{1}{*}{$t$} & 
    NLO & 169&820 & -1&3728 & 1&167 & 151&9990 & -1&3468 & 1&135 & 155&3376 & -1&3388 & 1&124 & 155&844 & -1&3288 & 1&108 \\
    \cmidrule{3-28}
    & & \multirow{1}{*}{$s$} & 
    NLO & 7&33867 & -3&6827 & 6&997 & 6&85043 & -3&6681 & 6&954 & 6&83761 & -3&6575 & 6&924 & 6&84724 & -3&6505 & 6&906 \\
    \cmidrule{3-28}
    & & \multirow{1}{*}{$\Wboson\topquark$} & 
    NLO & 33&2384 & -2&7747 & 4&46 & 31&7690 & -2&6608 & 3&795 & 34&3656 & -2&6566 & 3&785 & 34&6014 & -2&6496 & 3&791 \\
    \midrule
    \multirow{6}{*}{$\antitop$} & 
    \multirow{3}{*}{\begin{sideways}$\sqrt{s} = \unit[8]{\TeV}$\end{sideways}} & 
    \multirow{1}{*}{$t$} & 
    NLO & 29&8380 & -1&7837 & 1&785 & 28&6143 & -1&6984 & 1&655 & 30&786 & -1&6977 & 1&650 & 30&3703 & -1&6831 & 1&624 \\
    \cmidrule{3-28}
    & & \multirow{1}{*}{$s$} & 
    NLO & 1&88127 & -4&1505 & 8&396 & 1&82204 & -4&1082 & 8&266 & 1&89441 & -4&929 & 8&215 & 1&89198 & -4&774 & 8&173 \\
    \cmidrule{3-28}
    & & \multirow{1}{*}{$\Wboson\topquark$} & 
    NLO & 7&94316 & -3&1945 & 5&65 & 8&29082 & -3&3 & 4&593 & 9&3557 & -3&15 & 4&593 & 9&7282 & -3&128 & 4&639 \\
    \cmidrule{2-28}
    & \multirow{3}{*}{\begin{sideways}$\sqrt{s} = \unit[14]{\TeV}$\end{sideways}} & 
    \multirow{1}{*}{$t$} & 
    NLO & 97&1989 & -1&4759 & 1&314 & 90&3780 & -1&4234 & 1&244 & 94&5101 & -1&4187 & 1&235 & 94&8198 & -1&4029 & 1&211 \\
    \cmidrule{3-28}
    & & \multirow{1}{*}{$s$} & 
    NLO & 4&46678 & -3&8838 & 7&590 & 4&26563 & -3&8531 & 7&499 & 4&39473 & -3&8378 & 7&451 & 4&36731 & -3&8230 & 7&414 \\
    \cmidrule{3-28}
    & & \multirow{1}{*}{$\Wboson\topquark$} & 
    NLO & 33&537 & -2&7755 & 4&48 & 31&5711 & -2&6616 & 3&797 & 34&1677 & -2&6573 & 3&786 & 34&4158 & -2&6503 & 3&792 \\
    \bottomrule
  \end{tabular}
  \captionsetup{margin={2.em,0pt}} 
  \caption{Coefficients for fast cross section evaluations using 
    Eq.~\eqref{eq:mtapprox}. All cross section values are given in
    pb. We use always NLO PDFs.
    The
    reference top-quark mass $\bar\mt$ used here is the current world average
    of $\unit[173.5]{\GeV}$\ \cite{Beringer:1900zz}.}
  \label{tab:mtcoeff}
\end{sidewaystable}

\section{Example of \Hathor usage}
\label{sec:app-example}
\begin{verbatim}
#include "Hathor.h"
#include <iostream>

using namespace std;

int main() {
  Lhapdf pdf("CT10nlo");
  HathorSgTopT hathor(pdf);
  hathor.setColliderType(SgTop::PP);
  hathor.setSqrtShad(14000.);
  hathor.setScheme(SgTop::LO | SgTop::NLO);
  hathor.setPrecision(SgTop::LOW);

  double mt = 173., muf = mt, mur = mt;
  hathor.getXsection(mt, mur, muf);

  double val, err;
  hathor.getResult(0, val, err);
  cout << "cross section = (" << val << " +/- " << err << ") pb" << endl;

  return 0;
}
\end{verbatim}

\providecommand{\href}[2]{#2}\begingroup\raggedright\endgroup

\end{document}